\documentclass[10pt,twocolumn]{article}
\usepackage{times}

\usepackage[english]{babel}
\usepackage{blindtext}

\usepackage{amsmath,amssymb,amsfonts}
\usepackage{graphicx}
\usepackage{textcomp}
\usepackage{xcolor}

\usepackage{url}

\usepackage{breakurl}
\usepackage[breaklinks]{hyperref}

\usepackage{comment}
\usepackage{booktabs}
\usepackage{color}
\usepackage{xcolor}
\usepackage{xspace}
\usepackage{floatflt}
\usepackage{fancyhdr}
\usepackage{graphicx}
\usepackage{multirow}
\usepackage{subfig}
\usepackage{epstopdf}
\usepackage{float}
\usepackage{dblfloatfix}
\usepackage{amsmath}
\usepackage{amsthm}
\usepackage{amssymb}
\usepackage{enumitem}
\usepackage{outlines}

\usepackage{algorithm,algpseudocode,xcolor}
\usepackage[justification=centering]{caption}
\algdef{SE}[DOWHILE]{Do}{doWhile}{\algorithmicdo}[1]{\algorithmicwhile\ #1}%
\algnewcommand\algorithmicforeach{\textbf{for each}}
\algdef{S}[FOR]{ForEach}[1]{\algorithmicforeach\ #1\ \algorithmicdo}
\algblock{ParFor}{EndParFor}
\algnewcommand\algorithmicparfor{\textbf{parfor}}
\algnewcommand\algorithmicpardo{\textbf{do}}
\algnewcommand\algorithmicendparfor{\textbf{end\ parfor}}
\algrenewtext{ParFor}[1]{\algorithmicparfor\ #1\ \algorithmicpardo}
\algrenewtext{EndParFor}{\algorithmicendparfor}
\algtext*{EndIf}

\makeatletter
\def\therule{\makebox[\algorithmicindent][l]{\hspace*{.5em}\vrule height .75\baselineskip depth .25\baselineskip}}%

\newtoks\therules
\therules={}
\def\appendto#1#2{\expandafter#1\expandafter{\the#1#2}}
\def\gobblefirst#1{
  #1\expandafter\expandafter\expandafter{\expandafter\@gobble\the#1}}%
\def\LState{\State\unskip\the\therules}
\def\pushindent{\appendto\therules\therule}%
\def\popindent{\gobblefirst\therules}%
\def\printindent{\unskip\the\therules}%
\def\printandpush{\printindent\pushindent}%
\def\popandprint{\popindent\printindent}%

\algdef{SE}[WHILE]{While}{EndWhile}[1]
  {\printandpush\algorithmicwhile\ #1\ \algorithmicdo}
  {\popandprint\algorithmicend\ \algorithmicwhile}%

\algdef{SE}[DOWHILE]{Do}{doWhile}
  {\printandpush\algorithmicdo}[1]
  {\popandprint\algorithmicwhile\ #1}%
  
\algdef{SE}[FOR]{For}{EndFor}[1]
  {\printandpush\algorithmicfor\ #1\ \algorithmicdo}
  {\popandprint\algorithmicend\ \algorithmicfor}%
\algdef{S}[FOR]{ForAll}[1]
  {\printindent\algorithmicforall\ #1\ \algorithmicdo}%
\algdef{SE}[LOOP]{Loop}{EndLoop}
  {\printandpush\algorithmicloop}
  {\popandprint\algorithmicend\ \algorithmicloop}%
\algdef{SE}[REPEAT]{Repeat}{Until}
  {\printandpush\algorithmicrepeat}[1]
  {\popandprint\algorithmicuntil\ #1}%
\algdef{SE}[IF]{If}{EndIf}[1]
  {\printandpush\algorithmicif\ #1\ \algorithmicthen}
  {\popandprint\algorithmicend\ \algorithmicif}%
\algdef{C}[IF]{IF}{ElsIf}[1]
  {\popandprint\pushindent\algorithmicelse\ \algorithmicif\ #1\ \algorithmicthen}%
\algdef{Ce}[ELSE]{IF}{Else}{EndIf}
  {\popandprint\pushindent\algorithmicelse}%
\algdef{SE}[PROCEDURE]{Procedure}{EndProcedure}[2]
   {\printandpush\algorithmicprocedure\ \textproc{#1}\ifthenelse{\equal{#2}{}}{}{(#2)}}%
   {\popandprint\algorithmicend\ \algorithmicprocedure}%
\algdef{SE}[FUNCTION]{Function}{EndFunction}[2]
   {\printandpush\algorithmicfunction\ \textproc{#1}\ifthenelse{\equal{#2}{}}{}{(#2)}}%
   {\popandprint\algorithmicend\ \algorithmicfunction}%
\makeatother

\usepackage{mathtools}

\newcommand {\ie} {{\em i.e., }}
\newcommand {\eg} {{\em e.g., }}
\newcommand {\beq} {\begin{equation}}
\newcommand {\eeq} {\end{equation}}
\newcommand {\bequn} {\begin{equation*}}
\newcommand {\eequn} {\end{equation*}}
\newcommand {\bear} {\begin{eqnarray}}
\newcommand {\eear} {\end{eqnarray}}
\newcommand {\bearun} {\begin{eqnarray*}}
\newcommand {\eearun} {\end{eqnarray*}}
\newcommand {\fig}[1]{Fig.~\ref{#1}}
\newcommand{\figs}[2]{Figs.~\ref{#1} and~\ref{#2}}
\newcommand {\Eqref}[1]{Eq.~(\ref{#1})}

\newtheorem{theorem}{Theorem}
\newtheorem{lemma}[theorem]{Lemma}

\newtheorem{proposition}[theorem]{Proposition}

\begin{document}


\title{Impact of Distributed Rate Limiting on Load Distribution \\in a Latency-sensitive Messaging Service}

\author{Chong Li, Jiangnan Liu, Chenyang Lu, Roch Guerin, Christopher D. Gill \\
		    Washington University in St. Louis\\
          E-mail: \{chong.li, liu433, guerin, lu, cdgill\}@wustl.edu}

\date{}
\maketitle



\begin{abstract}

The cloud's flexibility and promise of seamless auto-scaling
notwithstanding, its ability to meet service level objectives (SLOs)
typically calls for some form of control in resources usage.  This
seemingly traditional problem gives rise to new challenges in a cloud
setting, and in particular a subtle yet significant trade-off
involving load-distribution decisions (the distribution of workload
across available cloud resources to optimize performance), and rate
control (the capping of individual workloads to prevent global
over-commitment). The paper investigates this trade-off through the
design and implementation of a real-time messaging system motivated by
IoT applications, and demonstrates a solution capable of realizing an
effective compromise.  The paper's contributions are in both
explicating the source of this trade-off, and in demonstrating a
possible solution.
\end{abstract}

\section{Background \& Motivations}
\label{sec:intro}

\subsection{Background}
\label{sec:background}
The cloud and its many ``*aaS'' instantiations~\cite{hoefer10} has
ushered in a new era of access to computations, and this has in turn
enabled an explosion in distributed applications, in particular in the
Internet-of-Things (IoT) space~\cite{botta16}.

IoT applications commonly involve a large volume of data generated
across many sources (sensors) distributed over geographically diverse
locations, and that need to be processed and often acted upon in a
timely manner, \eg for actuation purpose.  Consequently, they require
effective data transfer and processing solutions. The
  combination of the cloud's inherent computational flexibility and
  scalable communication platforms are what makes it an attractive
  platform for IoT
  applications~\cite{aws_iot,azure_iot,iotcase}. This has led to the
development of communication platforms such as Microsoft Azure Service
Bus and Amazon AWS IoT. Those platforms are based on a
  publish/subscribe (pub/sub) paradigm, which lets a large number of
  senders and receivers connect without the need for a complex mesh of
  one-to-one connections.

\begin{figure}[h]
\setlength{\belowcaptionskip}{-1pt}
\vspace{-0.1in}
\center
\includegraphics[width=\columnwidth]{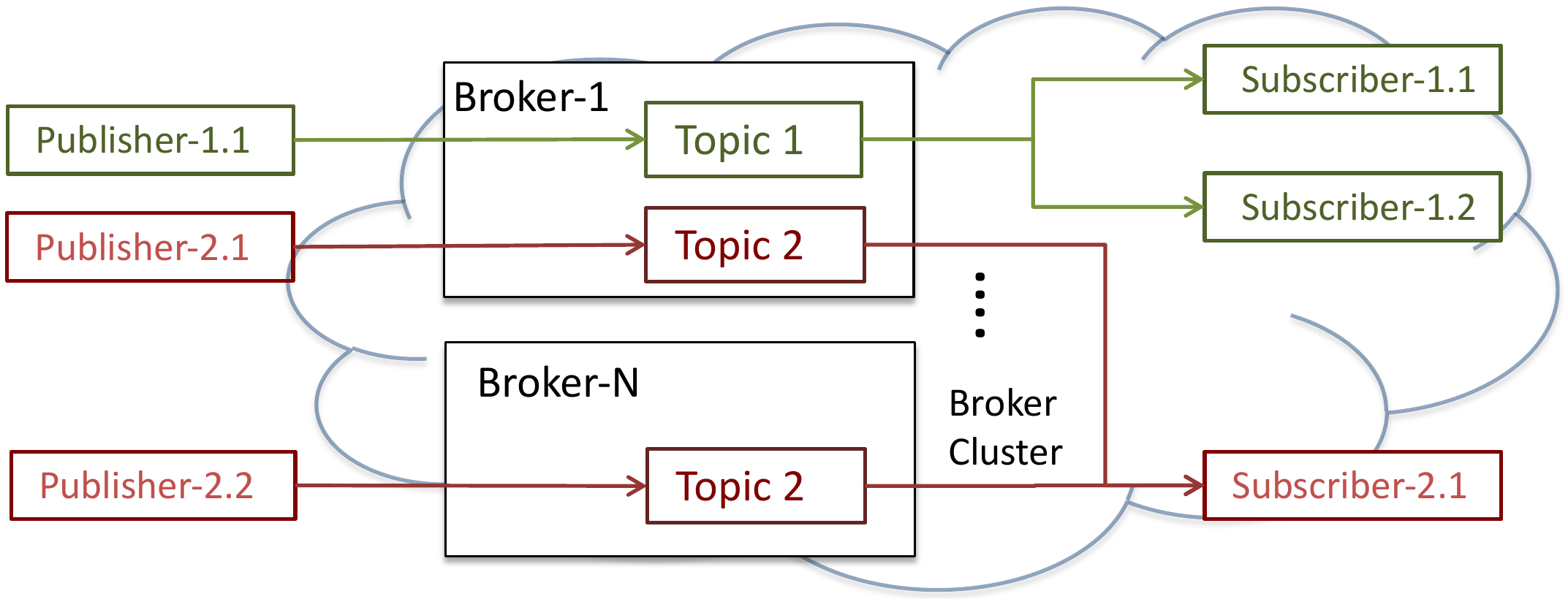}
\caption{Pub/sub messaging platform structure}
\vspace{-0.1in}
\label{pubsub}
\end{figure}

\fig{pubsub} illustrates the typical architecture for such a system,
with topics as the abstraction used to connect publishers (senders)
and subscribers (receivers).  Message brokers mediate between
publishers and subscribers by receiving, queueing and forwarding
messages for different topics. Publishers publish messages to a broker
\emph{for} a given topic, with subscribers subscribing to brokers to
receive messages \emph{from} that topic.  Scalability is realized by
having multiple brokers across which to distribute the
workload~\cite{topic_part_a,topic_part_k}, both from different topics
as well as for individual topics with a heavy message load, \eg
Topic~$2$ in \fig{pubsub}. This enables rapid access to additional
capacity when needed.

As with any shared resource, the workload of message
  brokers needs to be controlled to ensure that service level
  objectives (SLOs) are met. This is particularly important for IoT
applications that require timely delivery (and processing) of their
data.  If an application/topic was to (accidentally or intentionally)
misbehave and generate a much higher message load than anticipated, it
could overwhelm the platform resources (CPU and memory), and in turn
affect the SLOs of other topics. A standard approach
  to address this issue is to \emph{rate limit} the message volume of
  each topic.  Rate limiting is used in several public cloud
platforms and commonly implemented through a software API
gateway~\cite{aws_gateway,azure_gateway}.

In practice, the rate limiting mechanism is in the form of a
\emph{token
  bucket}~\cite{yahoo,barath,qjump,silo,aws_gateway,wcompactor}, where
each message requires a token before it can be processed by its
broker.  In the absence of tokens, an arriving message is deemed non-conformant and waits for
one (alternatively, it can be dropped).  A token bucket is
typically specified by two parameters, $(r,b)$, as part of the topic's SLO.
The parameter $r$ gives the rate at which tokens are generated, and
therefore bounds the topic's long-term message rate. The parameter $b$ indicates the maximum number of tokens the topic
can accumulate and bounds the maximum message burst it can
send without incurring an access delay (waiting for
tokens).  Their combination specifies the workload
``envelope''~\cite{mao06} that the rate controller enforces.

\subsection{A Motivating Example}
\label{sec:example}
Consider an intelligent transportation
  system~\cite{scats} that has to process vehicle volume data
acquired from tens of thousands of sensors distributed over an urban
region, and respond within a second or less~\cite{scats_traffic} to
ensure proper control of traffic signals.  In this system, the sensors
serve as publishers of information and cloud servers responsible for
processing that information as subscribers to the messaging service\footnote{In
closing the resulting control loop, actuation signals coming from the cloud servers,
now serving as publishers, are sent back, with traffic lights the 
corresponding subscribers.}.
The need for timely responses calls for provisioning the messaging
infrastructure to meet the system's SLO, typically in the form of
tail latency guarantee for message delivery (from publishers to
subscribers).

Because
multiple
topics share the same brokers, meeting SLOs calls for (rate) limiting
the users' message workload. Hence, the volume of messages our intelligent
transportation system sensors generate is first profiled, with this
profile used to configure its rate limiter, \ie to ensure little to no
access delay as long as it conforms to the corresponding traffic
envelope\footnote{In practice, the rate limiter is often configured
  with some ``margin'' to account for possible deviations from the
  original profile.}.

Implementing the rate limiting functionality at a single gateway, as is commonly done, has obvious limitations when it comes to scalability.
Furthermore, as rate limiting is typically in terms of application
data units, \eg messages, a gateway introduces an
additional application ``hop'' as it must reconstruct
(from TCP or UDP packets) application data units to (rate) control how many it lets in.  This extra application hop adds
latency that can be particularly detrimental to real-time
applications. These disadvantages have been acknowledged before and
have motivated the exploration of distributed rate limiting (DRL)
solutions~\cite{yahoo,tyk,barath,stanojevic09a,stanojevic09b}.

A DRL system involves multiple rate limiters, each
associated with a different resource to which an application has
been assigned, \eg brokers.  Those rate limiters then
collaborate to ensure that the aggregate traffic they allow
conforms to the same overall $(r,b)$ workload envelope as a
centralized limiter.  Solutions differ in the level of ``collaboration'' that exists between sub-token buckets.  At one extreme, sub-token buckets share (and update) a common state that they use to determine the conformance of messages.  This emulates centralized decisions while leaving enforcement distributed, but can incur a high communication overhead for state updates and queries by individual sub-token buckets. This is especially so in the presence of highly dynamic workloads.  At the other extreme, the original $(r,b)$ token bucket is statically split into sub-token buckets $(r_l,b_l)$, where $\sum r_l=r$ and $\sum b_l=b$ to preserve overall conformance with the original token bucket. This eliminates all communication overhead, and therefore ensures scalability independent of workload dynamics.  

This latter option is the one we focus on in this paper in spite of the fact that, as we shall see, it can introduce an additional access delay that we term the \emph{DRL penalty.}  Our goals are to explicate the origin of this penalty, and 
with this knowledge in hand to then propose, implement, and evaluate
a possible solution to mitigate it.
Specifically, Sections~\ref{sec: design} to~\ref{sec: eva} report on the design, implementation, and
evaluation of a scalable real-time messaging platform (SRTM) that leverages this insight.  SRTM is built on
top of the NSQ open-source messaging middleware~\cite{nsq} and is
available for others to use~\cite{srtm}.
  

\section{Problem statement \& Goal}
\label{sec: bg}

The basic load distribution (LD) question we seek to answer is as
follows: 
\begin{quote}
{\bf LD}: Given a new topic with rate limiter $(r,b)$ and a set
  of message brokers with existing workloads, how should we distribute
  publishers of the new topic across brokers to ``best'' meet the
  topic's SLO (target latency)?
\end{quote}
where our definition of \emph{best} is in terms of an efficient use of
resources, \eg yielding a greater residual capacity for equal performance, 
or the ability to support a higher messaging workload.

\begin{figure}[t]
\setlength{\belowcaptionskip}{-1pt}
\center
\includegraphics[width=\columnwidth]{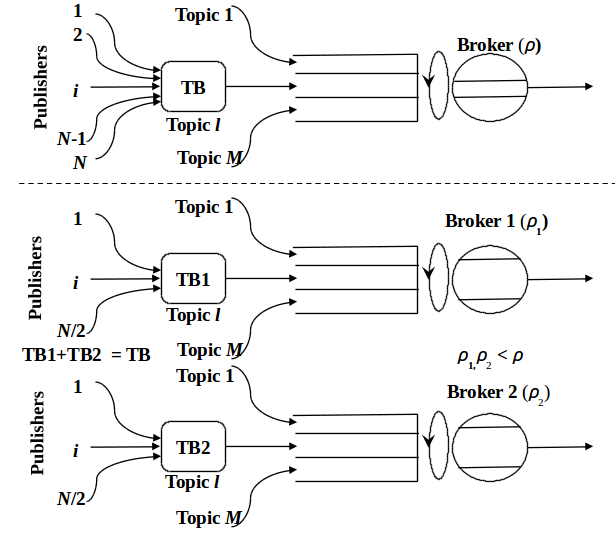}
\caption{Alternative DRL configurations}
\label{fig:2configs}
\end{figure}

This question is illustrated in \fig{fig:2configs} for a configuration
that involves a new topic that can either assign all its publishers to
a single broker with a resulting message processing utilization of
$\rho$ at the broker, or elect to split its publishers across two
brokers, each then with a message processing utilization of $\rho_1$
and $\rho_2$, respectively, where $\rho_1,\rho_2< \rho$. In the latter
case, the original topic's token bucket $(r,b)$ is split in two
sub-token buckets $(r_1,b_1)$ and $(r_2,b_2)$, one at each broker,
with, as stated earlier, parameters that verify $r=r_1+r_2$ and $b=b_1+b_2$.
Answering {\bf LD} then calls for identifying the configuration that
best meets the topic's performance (latency) goals.

This is a question that has been extensively investigated, even if
some care needs to be exercised in cases of servers with uneven
speeds, \eg see~\cite{rubinovitch85}.  As a general rule, access to
greater (message) processing capacity, \ie distributing the workload
across more processors, yields better raw performance (because of the
resulting lower load on individual processors), and this is embedded
in most load-balancing strategies.
The situation is different when latency is affected by \emph{both} the
message processing latency and the access delay that the rate limiting
function may introduce.  In particular and as we illustrate next,
splitting the rate limiting function has a negative impact on its
latency, so that the answer to {\bf LD} now involves a trade-off
between lowering message processing latency and increasing access
(rate limiting) latency.  Exploring this trade-off is a primary
motivation for this paper.


%

\subsection{Why splitting a token bucket is bad}
\label{sec:challenge}

In this section, we formally establish that distributing the rate limiting
function across multiple token buckets, \ie splitting them, has a
negative impact on the (access) delay that the token bucket itself
can introduce (what we termed the \emph{DRL penalty}).  In particular, even if message
and token rates are perfectly matched at the sub-token buckets across
brokers, the splitting alone ensures that messages experience a 
higher access (in the token bucket) delay.
As alluded to earlier, this increase in access delay
can all but offset the benefits of the greater message processing
capacity afforded by having access to multiple brokers.

Towards establishing this, we derive the stronger result that
splitting a token bucket into multiple sub-token buckets
always increases the cumulative rate limiting delay that messages
experience, \emph{irrespective} of how messages
are distributed across sub-token buckets.  The result is
formally stated in the next proposition whose proof can be found in
Appendix~\ref{app:2vs1}.

\begin{proposition}
Given a two-parameter token
  bucket $(r,b)$ and a general message arrival process where messages
  each require one token to exit the bucket, splitting this one-bucket
  system into multiple, say, $k$, sub-token buckets with parameters
  $(r_l,b_l)$ such that $r=\sum_{l=1}^kr_l$ and $b=\sum_{l=1}^kb_l$,
  can never improve the running sum of the message delays,
  irrespective of how messages are distributed to the $k$ sub-token
  buckets.  More generally, denoting as $S(t)$ and $S^{(k)}(t)$ the
  sum of the delays accrued by all messages up to time $t$ in the
  one-bucket and $k$-bucket systems, respectively, we have
\bequn
\label{eq:theo}
S^{(k)}(t)\geq S(t)\,, \,\forall t
\eequn
\label{theo:1vs2}
\end{proposition}
\vspace{-0.5cm}
While general, the result provides little
insight into its underlying cause.  To develop such insight, we explore
the special case of independent Poisson publishers for which the rate
limiting delay can be explicitly expressed.

Let $\lambda$ denote the aggregate (Poisson) message arrival rate of a
topic, and $(r,b)$ the parameters of its token bucket.  Under the
assumption of Poisson arrivals, the system behaves like a modified
M/D/1 queue~\cite{berger91,berger92} with a job arrival rate of
$\lambda$ and a service time of $1/r$ (the time needed to generate one
token), with messages delayed only when upon arrival the unfinished
work $U$ in the M/D/1 system exceeds~$b-1$.  The expected delay in an
$(r,b)$ token bucket is then of the form:
\bear
E[T_{TB}^{(1)}]
&=& \frac{1}{2r}\cdot\frac{P_{\mbox{M/D/1}}(U>b-1)}
                          {\left(1-\frac{\lambda}{r}\right)}
\label{eq:tb_md1}
\eear
where $P_{\mbox{M/D/1}}(U>b-1)$ can be computed as shown 
in~\cite[Section 15.1]{roberts96}
and captures the odds that a message is delayed in the token bucket
while $\frac{1}{2r\left(1-\lambda/r\right)}$ is the expected
delay of messages that have to wait for tokens.

Under Poisson arrivals, \Eqref{eq:tb_md1} still holds after (randomly) splitting
messages across, say, $k$ brokers, with corresponding message
arrival rates and sub-token bucket parameters that verify
$\sum_{l=1,\ldots,k}\lambda_l=\lambda$, $\sum_{l=1,\ldots,k}r_l=r$,
and $\sum_{l=1,\ldots,k}b_l=b$, where for ease of exposition, we
assume that $\lambda/r=\lambda_l/r_l, \forall\, l$, \ie message and token 
rates are perfectly matched,  

From \Eqref{eq:tb_md1}, we can readily identify the two factors that
contribute to the DRL penalty. 
Specifically,\\
{\bf (i)} $P_{\mbox{M/D/1}}(U_l>b_l-1)\geq P_{\mbox{M/D/1}}(U>b-1)$:
smaller buckets $(b_l\leq b)$ imply that messages are more likely to have to wait;\\
{\bf (ii)} $\frac{1}{2r_l}>\frac{1}{2r}$: with lower token rates $(r_l\leq r)$,
messages that have to wait (due to lack of tokens) wait longer.

Hence, \Eqref{eq:tb_md1} states that under the assumption of Poisson
arrivals, splitting the token bucket $k$-ways yields at least a
$k$-fold increase in access delay (\eg assuming $r_l=\frac{r}{k}, \forall\,
l$), and likely more (because\\ $P_{\mbox{M/D/1}}(U_l>b_l-1)\geq P_{\mbox{M/D/1}}(U>b-1)$).  

We note that the delay increase from the slower token rates $r_l$
is unavoidable and not dependent on the assumption of Poisson
arrivals. On the other hand, the fact that
$P_{\mbox{M/D/1}}(U_l>b_l-1)\geq P_{\mbox{M/D/1}}(U>b-1)$ is
explicitly dependent on the assumption of a Poisson 
process
hints at the possibility
that for different arrival processes this penalty may not always 
arise, or may be mitigated by
properly crafting the arrival process at each broker.

To better understand when and why this may be the case, consider a scenario 
where a token bucket $(r,b)$ is split in two equal sub-token buckets $(r/2,b/2)$.  Assume now that instead of independent Poisson publishers, publishers are synchronized, 
\ie generating messages at the same time to create an aggregate burst (batch Poisson).
Splitting publishers equally across sub-token buckets then also splits the burst
in the same proportion as the bucket size.  All other parameters, \eg load,
being the same, $P(U_2>b/2-1)\approx P(U>b-1)$, and the odds of messages being delayed are mostly unchanged\footnote{As mentioned earlier, the impact of lower token rates is still present.}.



In the next section, we discuss how we translate this intuition into a set of principles aimed at mitigating the DRL penalty while meeting topics' SLOs.

\section{SRTM Goals and Principles}
\label{sec: insight}

As reflected in {\bf LD}, SRTM seeks to offer a messaging service (for
IoT applications) that is both efficient in its use of cloud
resources and capable of enforcing latency guarantees (SLOs). A
common SLO is in the form of a tail latency guarantee, \eg a $99^{\mbox{th}}$ percentile latency below $1$~ms.  This
calls for both controlling the messaging workload that originates from
users (through rate limiting) and for determining how to best
distribute that workload across message processing resources
(brokers).  As discussed in the previous section, the difficulty lies
in the opposing effects of load distribution on message processing and
rate limiting latency, respectively.

A tongue-in-cheek restatement of the challenge faced by DRL decisions
would be ``to split, or not to split?'' Given the finding of
Proposition~\ref{theo:1vs2}, a natural guideline is to \emph{only
  split if you have to.}  In other words, distribute a topic's
publishers across the fewest brokers while ensuring that the resulting
message processing loads do not result in SLO violations.
Additionally, our intuition points to another postulate, namely,
\emph{if you split the load, split the burst}, at least to the extent
possible.  Specifically, publishers whose message transmission times
tend to be correlated, and therefore contribute to forming a burst,
should be assigned to different brokers.

Our approach to distributing publishers of a new topic to message
brokers builds on this insight.  Section~\ref{sec: design} provides
details on the resulting design, but we give next a brief overview 
and motivation for those
choices. Specifically, SRTM incorporates three principles:
\begin{enumerate}[itemsep=0pt]
\item \textit{Concentration}: Identify the smallest number of brokers
  needed to meet the new topic's SLO;
\item \textit{Max-min}: Maximize the minimum workload, and
  consequently token rate, assigned to any broker;
\item \textit{Correlation-awareness}: Assign publishers to brokers to
  minimize inter-publisher correlation, and consequently reduce the
  burstiness of the message arrival process at each broker as much as possible.
\end{enumerate}
\textit{Concentration} directly derives from
Proposition~\ref{theo:1vs2} that identifies that the DRL penalty is
unavoidable.  Hence, we seek to avoid or minimize it whenever
feasible.  In particular, as highlighted by \Eqref{eq:tb_md1}, the DRL
penalty can grow linearly and often super-linearly with the
number of sub-token buckets across which the workload is split.  Hence, 
it is natural to avoid splitting a topic as long as the broker's load
does not yield a processing latency that violates the topic's SLO.

\noindent
\textit{Max-min} is similarly inspired by \Eqref{eq:tb_md1} and the
fact that, irrespective of the arrival process, the access delay of
messages that experience a delay is inversely proportional to the
token rate.  Hence, keeping the minimum token rate across sub-token
buckets as high as possible is desirable for achieving tail latency guarantees.

\noindent
\textit{Correlation-awareness} seeks to select publishers so as to
decrease the ``burst'' of the
arrival process at each sub-token bucket in a manner that parallels
the decrease in the size of their respective bucket size.  We note
that decreasing arrival bursts should benefit both the DRL penalty and
the message processing delay in the brokers.

\section{SRTM Design}
\label{sec: design}

\begin{figure}[t]
\setlength{\belowcaptionskip}{-1pt}
\centering
\includegraphics[width=0.99\columnwidth]{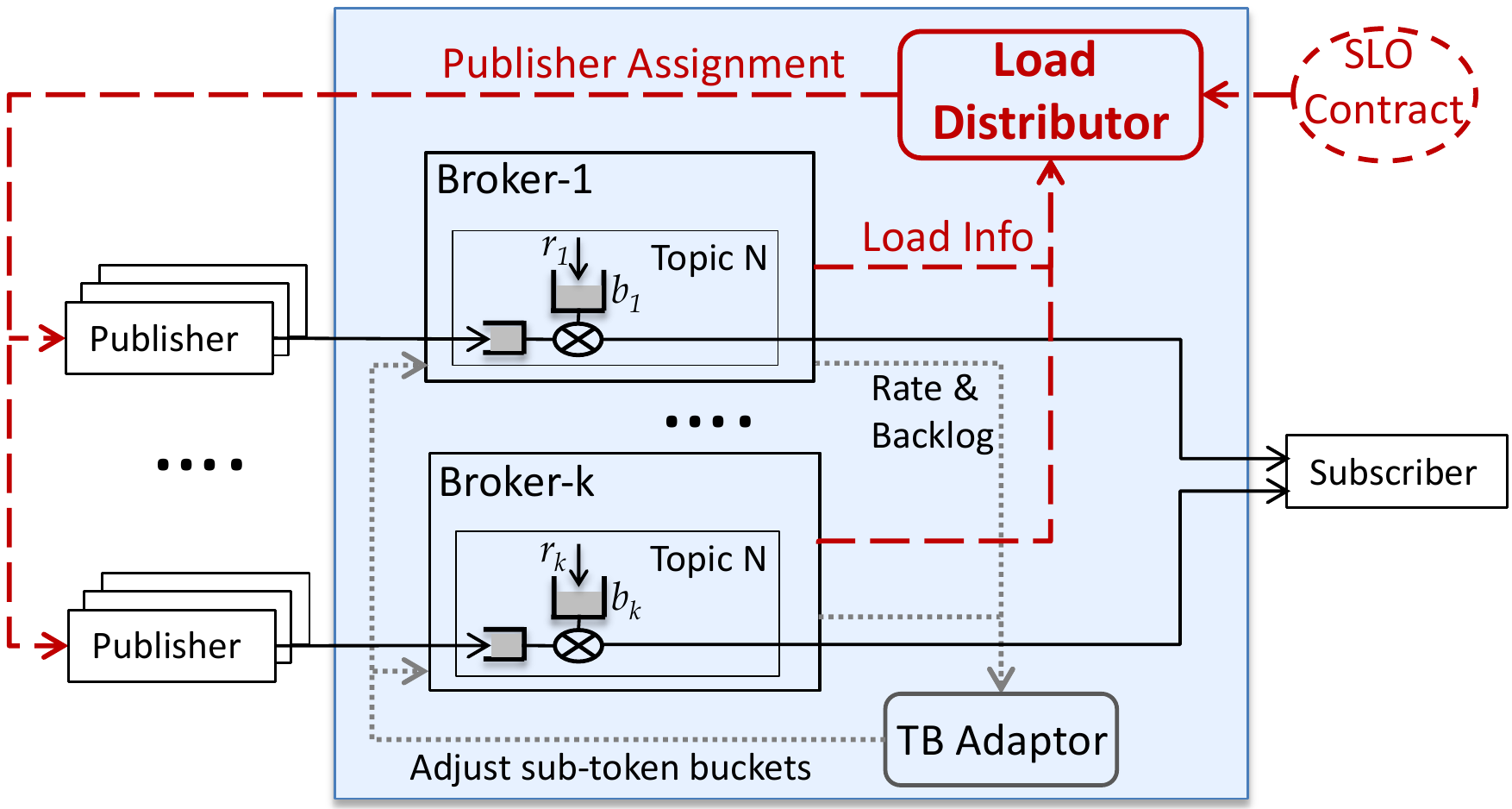}
\caption{SRTM Architecture Overview.\newline
The new topic (Topic N) is distributed across Broker-$1\sim k$
}
\label{SRTM}
\end{figure}

This section presents the design of SRTM, with \fig{SRTM} offering a
high-level overview of a typical configuration. 
The core component of SRTM is a \emph{Load Distributor} which is responsible for distributing a topic's publishers to brokers based on the topic's SLO and the brokers' existing workload.  In addition, a \emph{TB Adaptor} tracks the status of sub-token buckets at run time and triggers adjustments of their $(r_l,b_l)$ parameters in response to message traffic changes. In this section, we focus on the design of the SRTM \textit{Load Distributor,} as
it is the primary component responsible for realizing the principles
put forth in Section~\ref{sec: insight}.  Other components, including
the \emph{TB Adaptor}, are detailed in Section~\ref{sec: imp}.

\subsection{Design Challenges}
\label{sec:LB}

The Load Distributor is designed based on the three principles presented in Section~\ref{sec: insight}. However, realizing them calls for addressing two practical challenges. 

\textbf{Estimating Capacity.} In contrast to traditional load balancers SRTM does not seek to evenly distribute load across available resources. Instead, the \textit{Concentration} principle calls for determining the smallest number of brokers that can accommodate a new topic subject to its SLO (tail latency target). This, and to a lesser extent the \emph{Max-min} principle, is essentially an ``admission control'' problem, where the system keeps assigning publishers to a broker unless the topic's SLO is violated\footnote{We assume a common SLO across
topics.}.  Admission control requires an accurate estimation of the number of publishers a broker can handle subject to the SLO. 

In practice, estimating the broker capacity is challenging as message processing involves a set of inter-dependent and concurrent tasks. 
For example, the NSQ open-source messaging middleware~\cite{nsq} is implemented using the Go language (Golang)~\cite{golang}, which provides lightweight and scalable concurrency through Goroutines. This is well suited to IoT applications that involve large numbers of concurrent connections, and motivated our choice of NSQ as the basis for SRTM. However, modeling the behavior of the Goroutine runtime scheduler, including its reliance on a work-stealing strategy to exploit multicore systems, is non-trivial. Furthermore, depending on both the level of parallelism (number of publishers) of a topic and how publishers generate messages, performance bottlenecks migrate across NSQ components.  This makes a model-based approach mostly impractical and leads us to instead rely on a measurement-based approach~\cite{breslau00}.
Specifically, we use iterative measurements to discover how to best distribute a topic's publishers across brokers while meeting its latency target (see Section~\ref{sec:dist}).  

\textbf{Accounting for Correlation.} The other design challenge arises in realizing the \textit{correlation-awareness} principle. Specifically, based on \Eqref{eq:tb_md1}, given that a workload is to be split
across a number of brokers, SRTM's goal is to identify an assignment of publishers that results in the smallest possible increases in $P(U_l>b_l-1)$ across brokers.  This means crafting an arrival process at each broker that achieves this goal.
This is challenging as it requires precise temporal characterization of the workload generated by individual publishers. 

A reasonable option is to minimize the inter-arrival time variance at each broker. However, crafting such an outcome from individual publisher arrival processes is computationally complex.  Furthermore, variance only reflects global statistics of the arrival process at each broker, and so does not fully capture temporal correlation.  Hence, minimizing variance needs
not realize our goal of limiting increases in $P(U_l>b_l-1)$. Alternatives that rely on indices of dispersion~\cite{fendick89,gusella90,whitt19} are equally if not more complex, as is directly measuring temporal correlation.

Those challenges
lead us to instead rely on an altogether different alternative, namely, user-specified publisher correlation keys that reflect IoT application-level semantics (see Section~\ref{sec:correlation} for details).

\subsection{Iterative Workload Distribution}
\label{sec:dist}
When a new topic arrives, the Load Distributor is responsible for distributing its publishers among brokers to meet the topic's SLO. As discussed in 
Section~\ref{sec:LB}, it is challenging to accurately estimate the number of publishers a broker can accommodate subject to an SLO. To overcome this challenge, the Load Distributor employs an iterative process that distributes publishers across brokers as part of a measurement-based profiling phase. 

Each iteration of the load distribution process works as follows: (1) the Load Distributor first estimates the \textit{minimum} number of brokers ($k$) with enough available capacity to accommodate the new topic; 
(2) the topic's publishers are then assigned to the $k$ brokers in conformance
with the \textit{max-min} and \textit{correlation-awareness} principles; (3) after publishers have been assigned to the $k$ brokers, each broker runs an independent profiling phase (measuring latency) to validate whether it can accommodate its new workload without violating its SLO. 

After Step (3), each broker knows if it can handle its new workload or needs to shed some publishers to remain in compliance with its SLO. If there are no SLO violations, the load distribution process is deemed successful and ends. Otherwise, brokers whose SLO was violated determine how many publishers they need to shed, report this number to the Load Distributor, and mark themselves as ``full.'' The Load Distributor then starts a new iteration to distribute released publishers, possibly involving additional brokers. The process ends when all publishers are assigned to brokers that can accommodate them, or the system runs out of brokers.  When that happens, more brokers may be spawn by issuing a request to the cloud.

Next, we describe each step in more details, with the exception of the correlation-aware assignment of publishers, which is the subject of its own sub-section, Section~\ref{sec:correlation}. For simplicity, we first describe the process used in the first iteration that takes place when a new topic arrives. We then extend the discussion by describing how the variables used in each iteration are updated.
{\bf Step~(1)} reflects our \textit{concentration} goal and seeks to determine the minimum number of brokers needed to accommodate a new topic with a given aggregate message $rate$.  The
residual message processing capacity ($rcap$) of each broker is first estimated based on the difference between its maximum message processing capacity\footnote{This depends on the SLO and relies on a benchmarking phase.} ($mcap$), and its currently allocated message rate (the sum of the message $rate$s from topics already assigned to the broker). 
Brokers are then sorted in decreasing order of $rcap$ value to identify the \textbf{minimum} initial number $k$ of brokers to which to assign the new topic's workload (the smallest number of brokers such that the sum of their $rcap$ values exceeds the topic's message \emph{rate}). 

{\bf Step~(2)} determines how to best distribute the message workload of the new topic across those $k$ brokers according to the \textit{max-min} and \textit{correlation-awareness} principles. This involves computing for each broker a workload $quota$ it should be assigned.
This $quota$ is set to the minimum of the broker's residual capacity $rcap$, and its fair share $rate/k$ of the workload.
This maximizes the minimum assignment at each broker (the \textit{max-min} principle), unless limited by residual capacity. Once $quotas$ are set, brokers $rcap$ values are updated to reflect their new allocation, and publishers are assigned to brokers to realize those allocations in a \textit{correlation-aware} manner, which, as mentioned earlier, is described in Section~\ref{sec:correlation}. 

{\bf Step~(3)} acknowledges that estimates for $rcap$ only account for message rates, and therefore ignore many factors that affect performance, \eg arrival burstiness, concurrency (among connections from different publishers), interactions across workloads, etc.  Step (3), therefore, relies on a measurement-based approach to evaluate the \emph{actual} performance of each broker after Step~(2). 

Specifically, Step~(3) includes a profiling phase, which we call \textbf{online-fitting}, whose
goals are to (i) test whether each broker's SLO is still met after adding the new publishers (by measuring end-to-end message latency), and (ii) if it is not, determine how many publishers the broker needs to shed to return to compliance with the SLO. Under (ii), an iterative binary search\footnote{This is performed in parallel by all the brokers with violated SLOs.} is triggered to identify the how many publishers the broker can accommodate without violating its SLO.
Excess publishers are returned to the Load Distributor, with the broker marked as full.

This next iteration parallels the first one with updated variables. A new variable $ua\_rate$ records the aggregate unassigned message rate from excess publishers, while $k$ is increased to account for the smallest number of additional brokers needed to accommodate $ua\_rate$ in a manner that again conforms to the \emph{max-min} and \emph{correlation-awareness} principles.
Iterations continue until all publishers are assigned to brokers that can accommodate them, or the system runs out of brokers. 

In spite of its accuracy, we acknowledge that this iterative process has disadvantages. First, its measurement-based nature implies that convergence can take time.  We quantify this in Section~\ref{sec:time}.  Second, the SLO needs not be met while we iterate. This is unavoidable in any measurement-based approach and, as discussed in Section~\ref{sec:LB}, is a consequence of the difficulties in constructing a sufficiently general and accurate model.

\subsection{Correlation-aware Allocation}
\label{sec:correlation}

As pointed out in Section~\ref{sec:LB}, measuring correlation across a topic's publishers is challenging. As a result, we adopt a pragmatic approach where users provide correlation information based on application semantics commonly available in IoT settings. Our approach is inspired by the concept of partition keys used in stream processing in Azure IoT hub~\cite[p.~$123$]{basak17}, which allows users to explicitly identify correlated data streams to facilitate efficient stream processing by jointly considering streams within the same partition. Partitions that give rise to temporal correlation among publishers are common in many IoT deployments. For example, sensors in close proximity to each others will often trigger around the same time when detecting the same physical phenomenon. Publishers associated with those sensors, will then produce correlated message arrival patterns to the messaging system.  

We adapt the concept of partition keys in SRTM, and provide APIs to users that let them label correlated publishers with an identical \textbf{correlation group key}. 

Returning now to the end of Step~(2) of the workload distribution process, the Load Distributor seeks to assign publishers with the same correlation key to different brokers, thereby splitting the message ``bursts'' they jointly generate.  Specifically, given a set of correlation groups (as specified by the user) and a number of publishers to be assigned to a broker (based on $rcap$ and $rate/k$), the Load Distributor selects publishers from each correlation group in proportion to the number of publishers in the group.

\section{SRTM Implementation}
\label{sec: imp}

As mentioned earlier, 
SRTM is based on NSQ.
In this section, we first introduce NSQ's architecture, and then present our approach to adding SRTM's functionality together with some challenges we had to address. While discussions on implementation challenges are closely tied to NSQ, several of the issues we encountered are broadly applicable to systems that need to achieve predictable performance while relying on high concurrency platforms (for scalability) such as Go runtime.

\subsection{NSQ Architecture Overview}
\label{sec: nsq}

\begin{figure}[h]
\setlength{\belowcaptionskip}{-10pt}
\center
\includegraphics[width=0.45\textwidth]{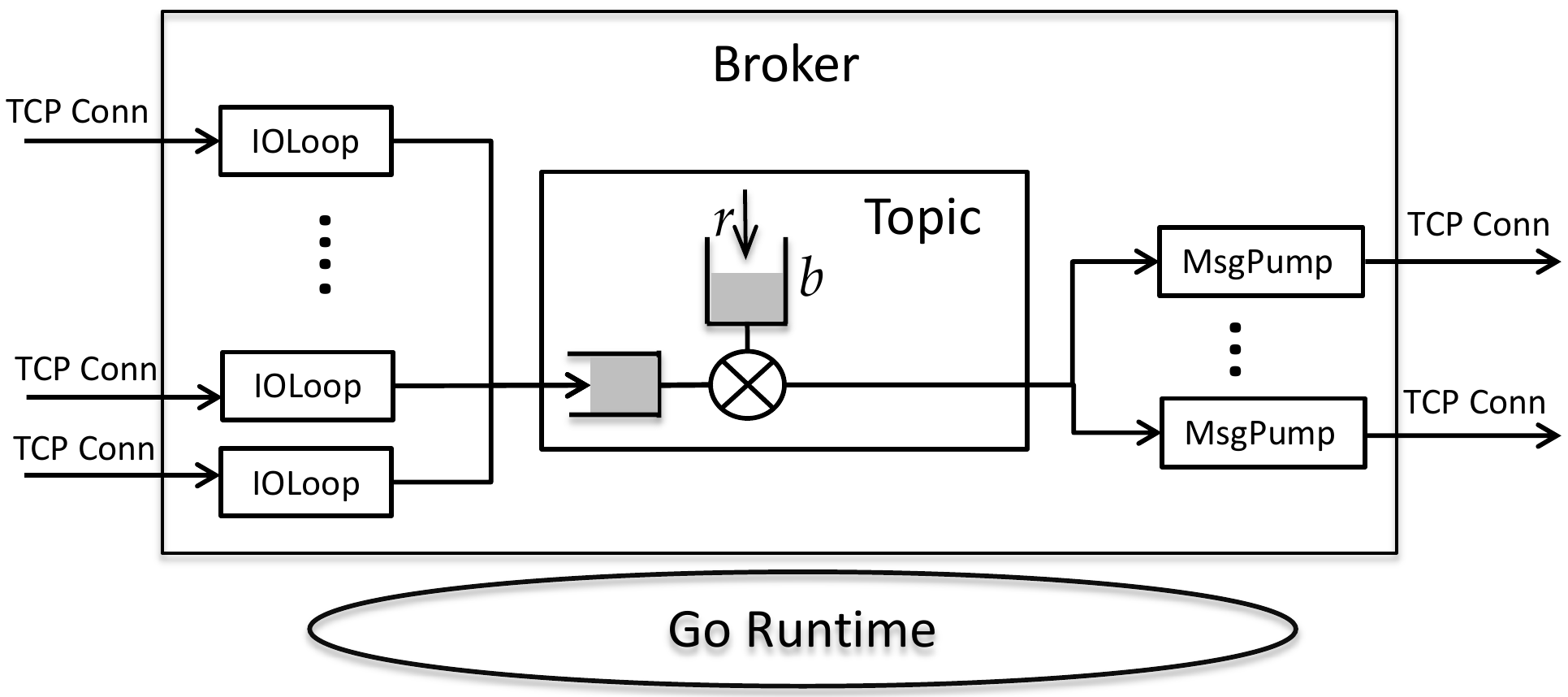}
\caption{Goroutines in a NSQ broker}
\label{goroutine}
\end{figure}

As shown in \fig{goroutine}, NSQ comprises a set of goroutines. Messages from publishers arrive on separate TCP connections each handled by an {\tt IOLoop} goroutine.  Messages from publishers of a given topic are then passed to another goroutine, {\tt Topic}. The {\tt Topic} goroutine has a dedicated buffer into which {\tt IOLoop} goroutines move messages. Once a {\tt Topic} goroutine is scheduled for execution, it runs to completion by pulling messages from its buffer and forwarding them to {\tt MsgPump} goroutines until the buffer becomes empty. The {\tt MsgPump} goroutines sends messages to subscribers through separate TCP connections. The Go runtime employs a work-stealing scheduler to schedule all the goroutines on multicore platforms. The lightweight goroutines and the scheduler are important to the scalability of NSQ that may need to handle a large number of publishers through a single broker. 

\subsection{Rate Limiting }
\label{sec: goroutine}

SRTM enforces rate limits through a token bucket that resides in the {\tt Topic} goroutine. At runtime, each {\tt Topic} goroutine keeps track of the state of its token bucket. When a {\tt Topic} goroutine is scheduled, it pull messages from its incoming message buffer only if its token bucket has tokens. Otherwise, the routine returns and messages wait in the message buffer until tokens become available.


While the logic of a token bucket is well understood, it is non-trivial to implement its precise temporal behavior in a high concurrency environment such as Go runtime.  Specifically, the token bucket determines whether a message is conformant based on its state (number of tokens) \emph{and} the message arrival time. The straightforward approach to measure arrival times is to timestamp messages when the {\tt IOLoop} goroutine reads them from the TCP receive buffer. However, because of the cooperative scheduling of goroutines, the times at which an {\tt IOLoop} goroutine is scheduled to read messages can significantly differ from when messages are first received in the TCP buffer. As the token bucket logic updates its token count based on how much time has elapsed since the message arrived, a late arrival timestamp can cause the token bucket to mistakenly delay messages that are actually conformant. Such scheduling delays can become significant when the number of goroutines in the system is large.  In particular, a topic with a large number of publishers, and consequently {\tt IOLoop} goroutines, might introduce large temporal errors in the token bucket's behavior.

To eliminate those errors,
SRTM employs TCP-layer timestamping.  The arrival of each TCP packet (sk\_buff) is recorded inside the Linux kernel (in the tcp\_v4\_rcv function). Whenever an {\tt IOLoop} reads data, the TCP arrival timestamps are also copied (as out-of-band data) to user space. As user data (messages) often do not map 1:1 to TCP packets (fragments), each {\tt IOLoop} maintains a mapping between received messages and TCP packets and assigns each message the timestamp from the correct TCP packet. As arrival timestamps at the TCP-layer are independent of goroutine scheduling, SRTM is able to enforce rate limits with high temporal accuracy.

\subsection{Handling Garbage Collection}
Garbage Collection (GC) in the Go runtime can have a significant impact on the tail latency of message processing in NSQ. When GC is triggered, Go runtime uses marker goroutines to mark memory allocations, which can consume up to 25\% of CPU time~\cite{gogc}. As a result, we found significant increases in the $99^{\mbox{th}}$ percentile in the message processing tail latency when GC is triggered.

As GC is usually triggered on demand in Go~\cite{gcpercent}, we can minimize GC invocations by reducing dynamic memory allocation and hence slowing the growth of the heap size. We developed a GC-friendly version of NSQ by replacing instances of dynamic memory usage with statically allocated memory. Specifically, we created a pre-allocated ring buffer for each {\tt IOLoop}, such that data read from the TCP socket is directly written into an existing slot in the ring buffer instead of having to request a dynamic memory allocation. In addition to on-demand GC, Go runtime forces GC if there is no GC in a $2$ minutes (by default) interval. We disabled this feature so that in SRTM, GC is solely triggered on demand.
Although these optimizations do not completely eliminate the impact of GC, they effectively mitigate it for the $99^{\mbox{th}}$ percentile tail latency in our experiments.

\subsection{Adapting to Workload Changes}
\label{sec: rebal}

As discussed in Section~\ref{sec:dist}, SRTM performs load distribution upon the arrival of a new topic. Traffic may, however, shift over time across brokers, while remaining conformant to the global $(r, b)$ traffic profile. The resulting mis-match between traffic and sub-token buckets can then produce significant DRL penalties. To avoid this, SRTM's static rate control configuration is augmented with a TB Adaptor that adjusts sub-token bucket parameters in response to ``long-term'' traffic shifts.

Specifically, every broker periodically reports to the TB Adaptor the rate and burst statistics of its topics.  In our implementation, this is based on a per-topic 10s history window that tracks the average message rate and maximum backlog over that window in 1s increments.  The TB Adaptor adjust token rates in proportion to average message rates, and token bucket sizes in proportion to maximum backlogs. To balance responsiveness and stability, SRTM administrators can adjust the length of the history window and the frequency of updates.

Note that the TB Adaptor only adjusts sub-token bucket parameters to avoid DRL penalties caused by shift in traffic among publishers. This does not address problems that may arise when a shift in traffic overloads a broker.  Handling such scenarios calls for the ability to migrate publishers among brokers.  Although this feature has been implemented (and is used in the profiling phase), its introduction as a runtime mechanism is left as future work.

\section{Evaluation}
\label{sec: eva}

This section presents an empirical evaluation of SRTM, and more precisely of the different design principles on which it relies. 
Section~\ref{sec: eva-drl} starts with experiments designed to illustrate the impact of DRL on tail latency, while Sections~\ref{sec: con} to~\ref{sec:corr} proceed with quantifying the relative benefits derived from each one of SRTM's three principles. This is realized through a progression of designs that incorporate SRTM's principles one at the time.  Finally, Section~\ref{sec:time} explores a more pragmatic aspect, namely, the amount of time SRTM takes to converge to a stable load distribution after the arrival of a new topic.

\begin{figure}[h]
\setlength{\belowcaptionskip}{-10pt}
\center
\includegraphics[width=\columnwidth]{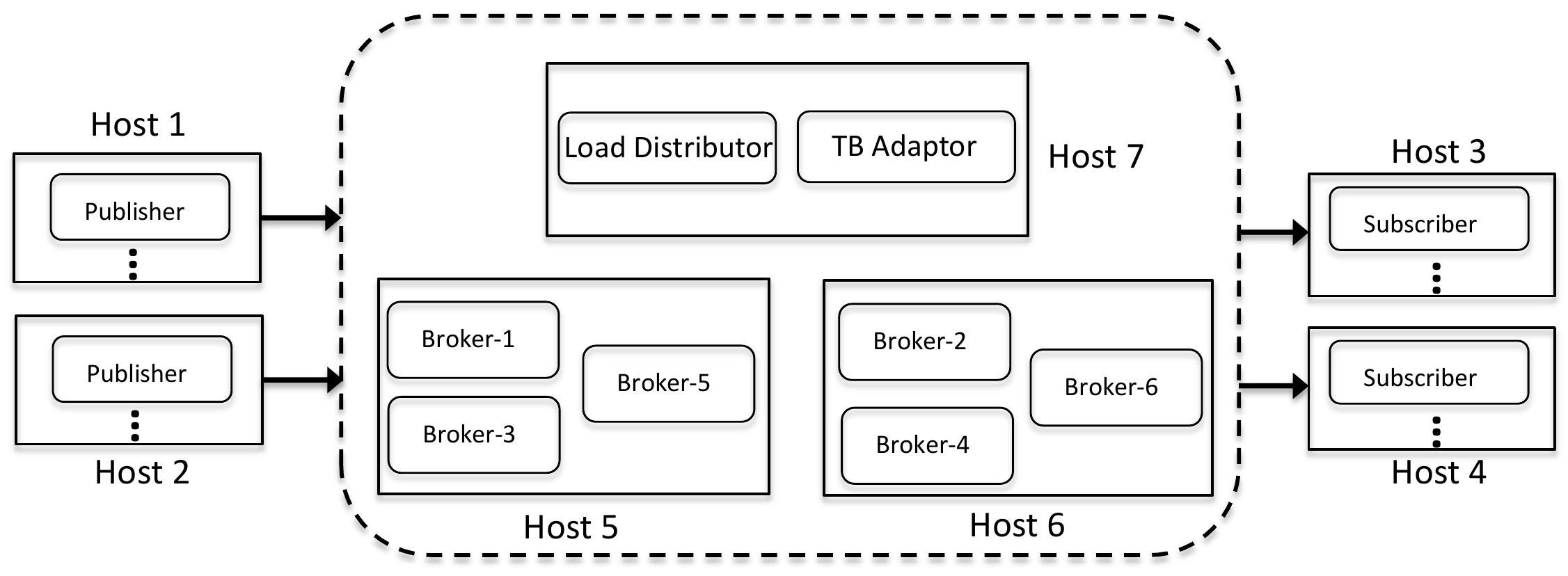}
\caption{Testbed Setup}
\label{fig:setup}
\end{figure}

\textbf{Testbed.} The evaluation is carried out on a testbed consisting of 7 physical hosts. \fig{fig:setup} offers an overview of the testbed. Hosts boast two 8-core Intel Xeon E5-2630 processors, 8~GB of memory, and are connected by 40~Gbps Ethernet links.  Hosts~$1,2$ and~$3,4$ are dedicated to publishers and subscribers, respectively. There are $6$ brokers in total, deployed over Hosts~$5$ and~$6$. Each broker has $2$ dedicated CPU cores, which is also the default CPU configuration of Amazon-MQ instances~\cite{awsmq}. The Load Distributor and TB Adaptor are deployed at Host~$7$. 

\textbf{Workload.} We generate a messaging workload that seeks to mimic IoT traffic. In a typical IoT setup, messages for a topic originate from multiple publishers, with publishers corresponding to a single sensor or a gateway aggregating a group of sensors. Messages can be \textit{time-triggered} or \textit{event-triggered}. 
Time-triggered messages are usually generated periodically. For example, intelligent transportation systems may require periodic traffic updates from sensors across a city~\cite{citypulse, scats}. 
Conversely, event-triggered messages are generated upon detecting specific environmental changes. For example, building thermostats may trigger when temperature drops below (or exceeds) a certain threshold.  

We emulate time-triggered traffic with publishers sending messages periodically.  Because we were unable to secure real-world traces of event-triggered messages, we approximated the resulting traffic using a Poisson process (randomly occurring events). Both periodic and Poisson publishers may generate messages in a \textit{batch}, with the \textit{batch size} determining the traffic burstiness. For example, a gateway controlling a group of sensors would generate a batch of messages if the sensors are controlled by a common timer or triggered by the same event.

We selected a $99^{\mbox{th}}$ percentile end-to-end message latency of $1ms$ as the SLO\footnote{It reflects the typical end-to-end latency in our testbed. A real-world deployment would likely have to account for larger network delays.} across all experiments.
Whenever a given design is unable to meet the $1ms$ SLO, we report the performance of the best configuration. Additionally, topics' token buckets were configured using a profiling phase that gathered a representative traffic trace. The trace was used to perform an offline simulation\footnote{For Poisson publishers, the approach behind \Eqref{eq:tb_md1} could instead be used.} of the token bucket performance using a token rate $r$ set $10\%$ higher than the topic's message rate, and searching for the smallest bucket size $b$ that ensured a $99^{\mbox{th}}$ percentile token bucket access delay of zero. 




\subsection{Illustrating the DRL Penalty}
\label{sec: eva-drl}

\begin{figure}
         \subfloat[1,000 publishers \label{single1}]{%
      \includegraphics[width=0.24\textwidth]{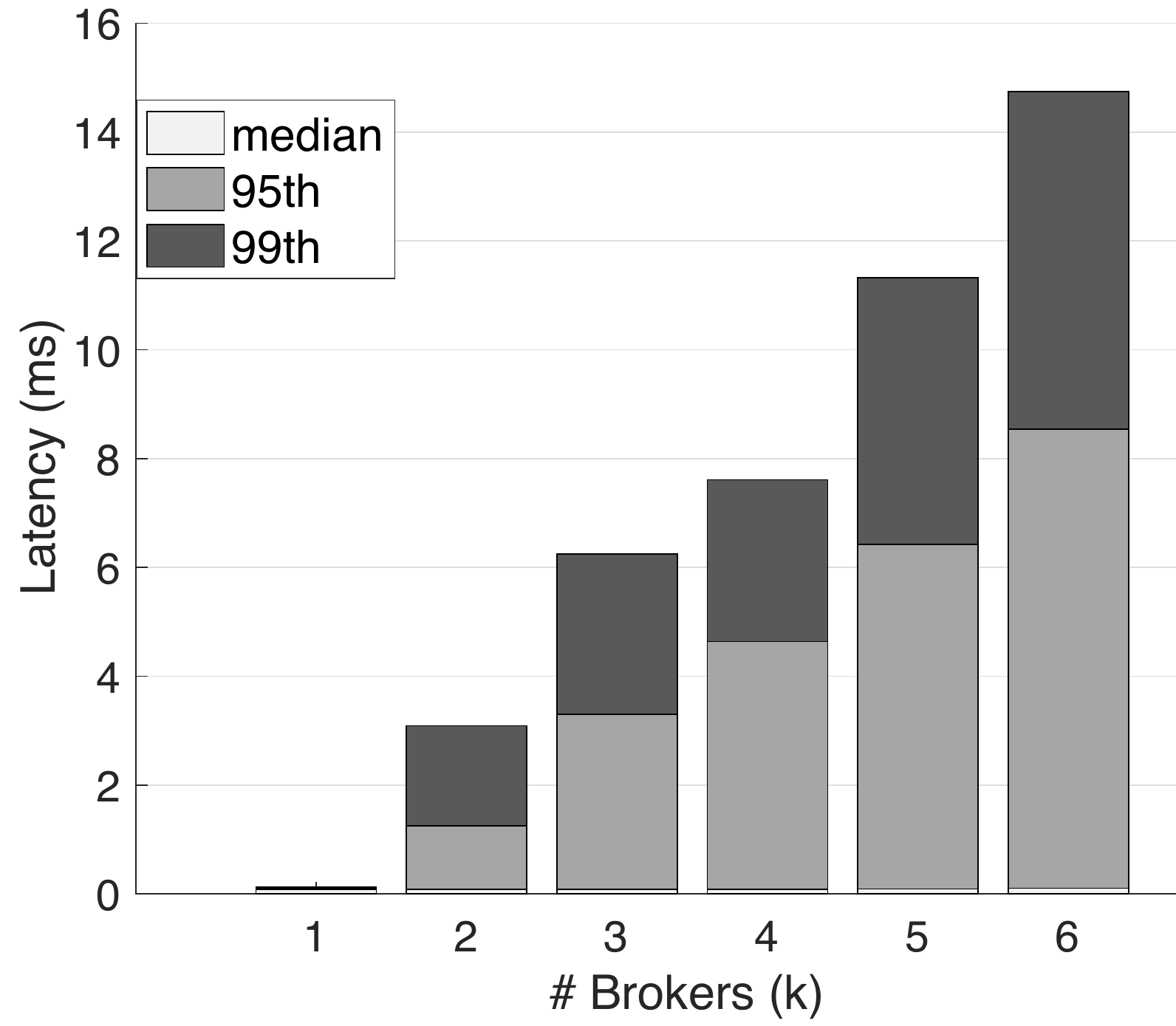}
    }
    ~
    \subfloat[6,000 publishers \label{single2}]{%
      \includegraphics[width=0.24\textwidth]{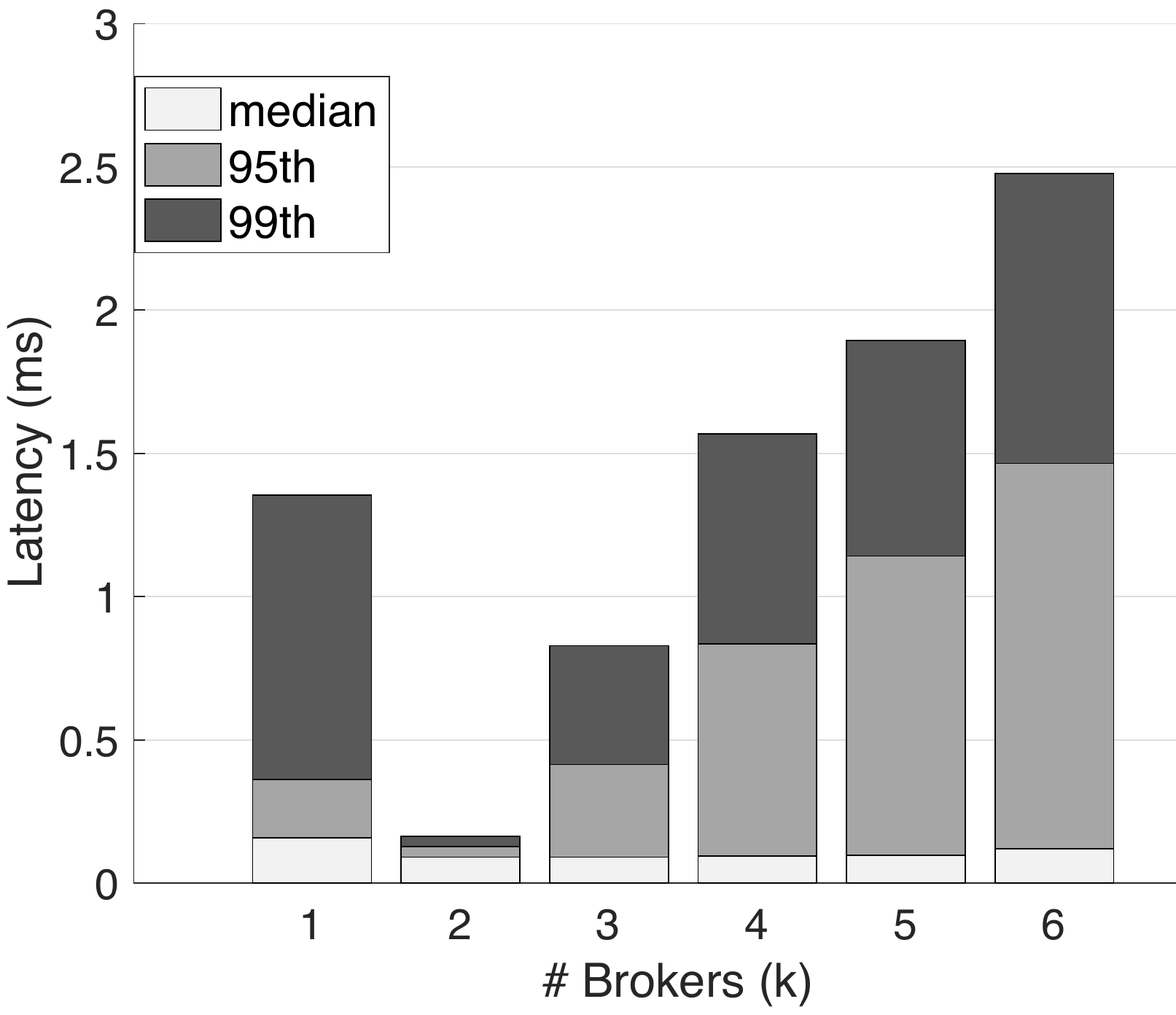}
    }
    \caption{DRL penalty for Poisson workloads\\ 
    (10 msg/s per publisher -- batch size of 1).}
    \label{single}
\end{figure}

This section explores the impact of the DRL penalty on tail latency through two experiments (see \fig{single}) that differ in their message workload. In both experiments, the publishers of a new topic have access to the $6$~brokers, which for consistency are initially all idle.  The first experiment illustrates the DRL penalty when the topic's load is low, \ie a single broker is nominally able to accommodate the topic.  The second experiment considers a heavier load for which a trade-off arises between access to more capacity when spreading the load across brokers vs.~the resulting increase in DRL penalty.  Both experiments rely on publishers that generate single messages (batch size of~$1$) according to a Poisson process of rate of $10~msg/s$.

The first experiment involves a topic with $1,000$ (Poisson) publishers (a workload of $10,000~msg/s$) that are distributed across 1 to 6 brokers, with the topic's token bucket correspondingly split among them. \fig{single1} shows the median, $95^{\mbox{th}}$, and $99^{\mbox{th}}$ percentile of the end-to-end latency for the 6 configurations.  It illustrates the increase of the $99^{\mbox{th}}$ percentile latency with the number of brokers across which the topic is split (it reaches $14.7ms$ for $k=6$). This is because in low load configurations, as is the case here, the benefit from access to more broker capacity is small and does not offset the DRL penalty.

The second experiment (\fig{single2}) is similar except that it now uses $6,000$~publishers. The higher message rate overloads a single broker, as illustrated by the case $k=1$ that exhibits a large tail latency. Splitting publishers between two brokers reduces the tail latency, as the decrease in message processing latency (from the lower broker load) exceeds the increase in DRL penalty (from splitting the token bucket). Distributing the topic across more brokers is, however, of no benefit, as the increase in DRL penalty again exceeds the decrease in message processing latency. This highlights the trade-off that \fig{fig:2configs} alluded to. 

Next, we compare SRTM against a series of baseline solutions that incrementally incorporate the design principles behind SRTM, namely, \textit{concentration}, \textit{max-min}, and \textit{correlation-awareness}. The experiments allow us to isolate the contributions of each principle, while assessing their overall impact when combined in SRTM. 

\begin{figure}
    \subfloat[Poisson, Batch=1, \label{com1-1}]{%
      \includegraphics[width=0.24\textwidth]{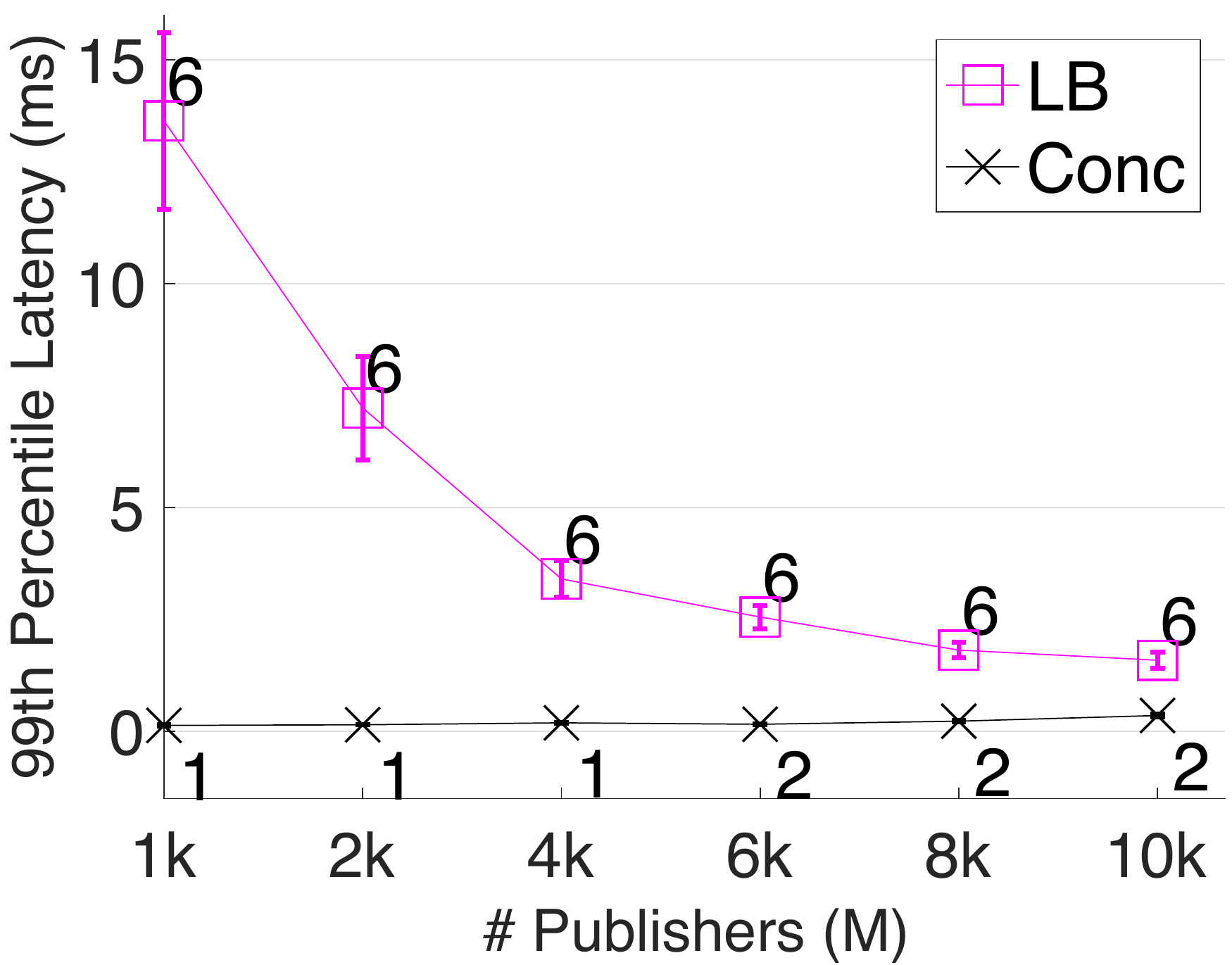}
    }
    ~
        \subfloat[Poisson, Batch=10, \label{com1-2}]{%
      \includegraphics[width=0.24\textwidth]{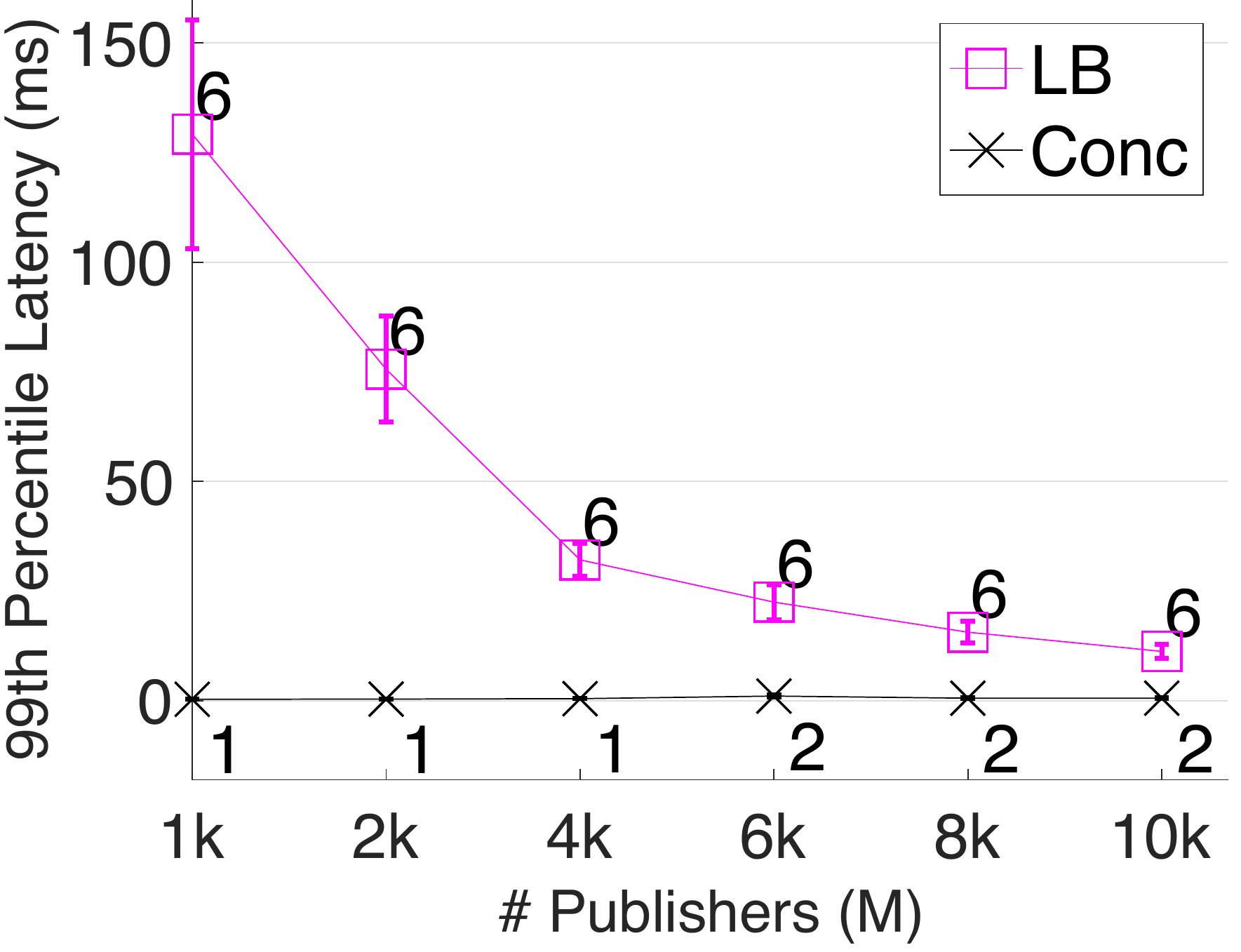}
    }
    ~
    \newline
    ~
     \subfloat[Periodic, Batch=1, \label{com1-3}]{%
      \includegraphics[width=0.24\textwidth]{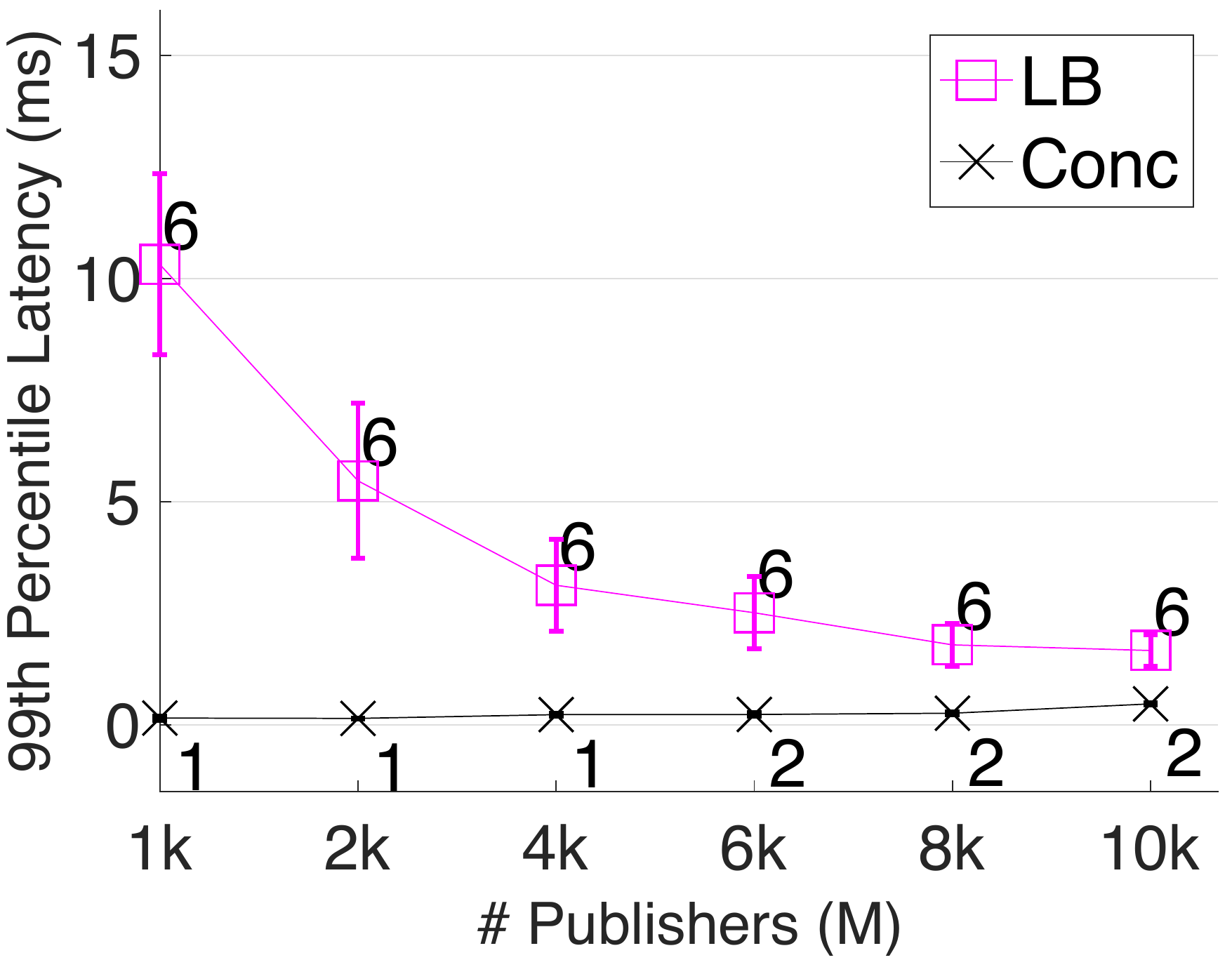}
    }
    ~
        \subfloat[Periodic, Batch=10, \label{com1-4}]{%
      \includegraphics[width=0.24\textwidth]{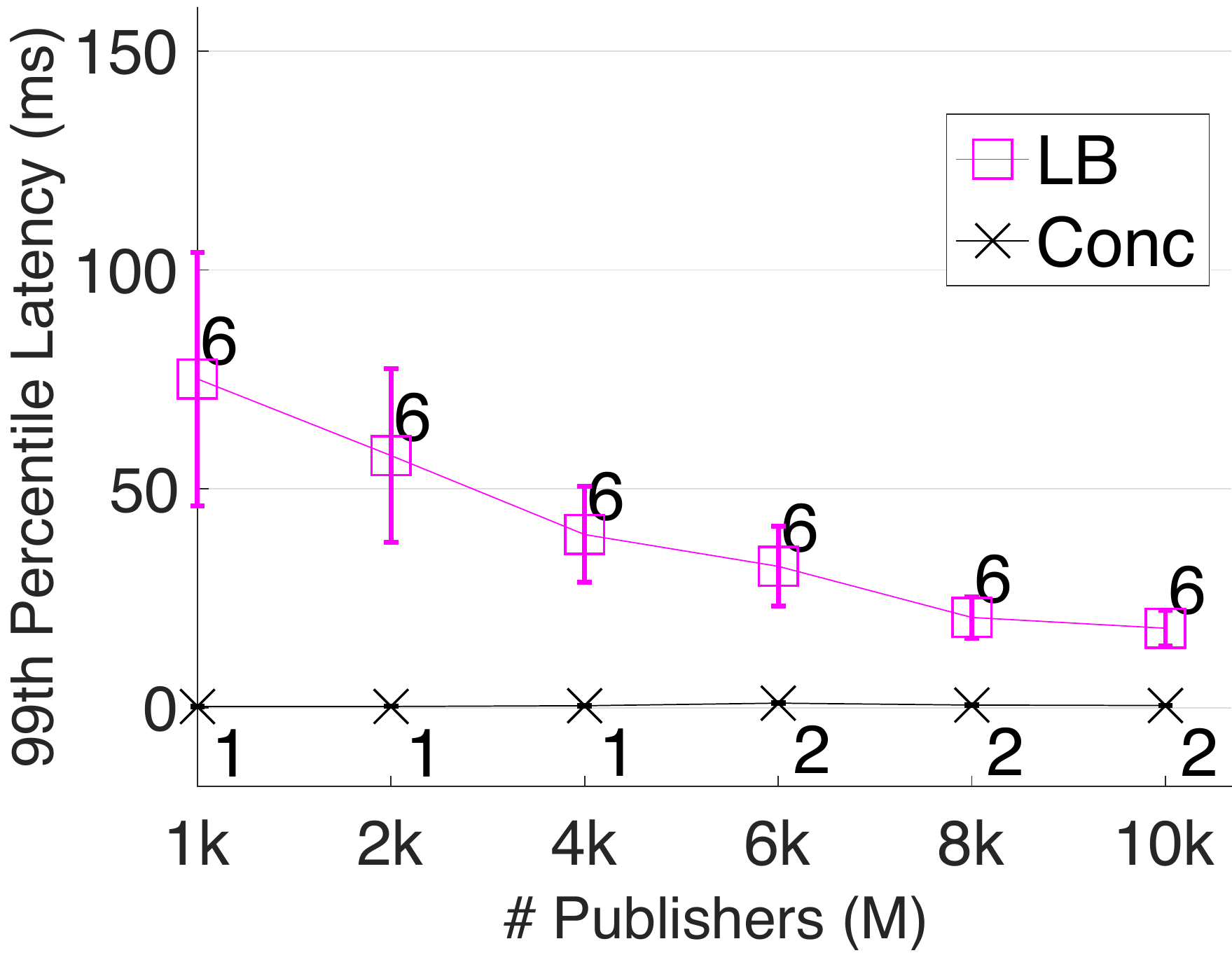}
    }
    ~
    \caption{Impact of Concentration on $99^{\mbox{th}}$ percentile Latency. (Batch: batch size per publisher; average message rate per publisher: 10 message/s.)}
    \label{com1}
\end{figure}

\begin{figure}
    \subfloat[Poisson, Batch=1\label{com2-1}]{%
      \includegraphics[width=0.24\textwidth]{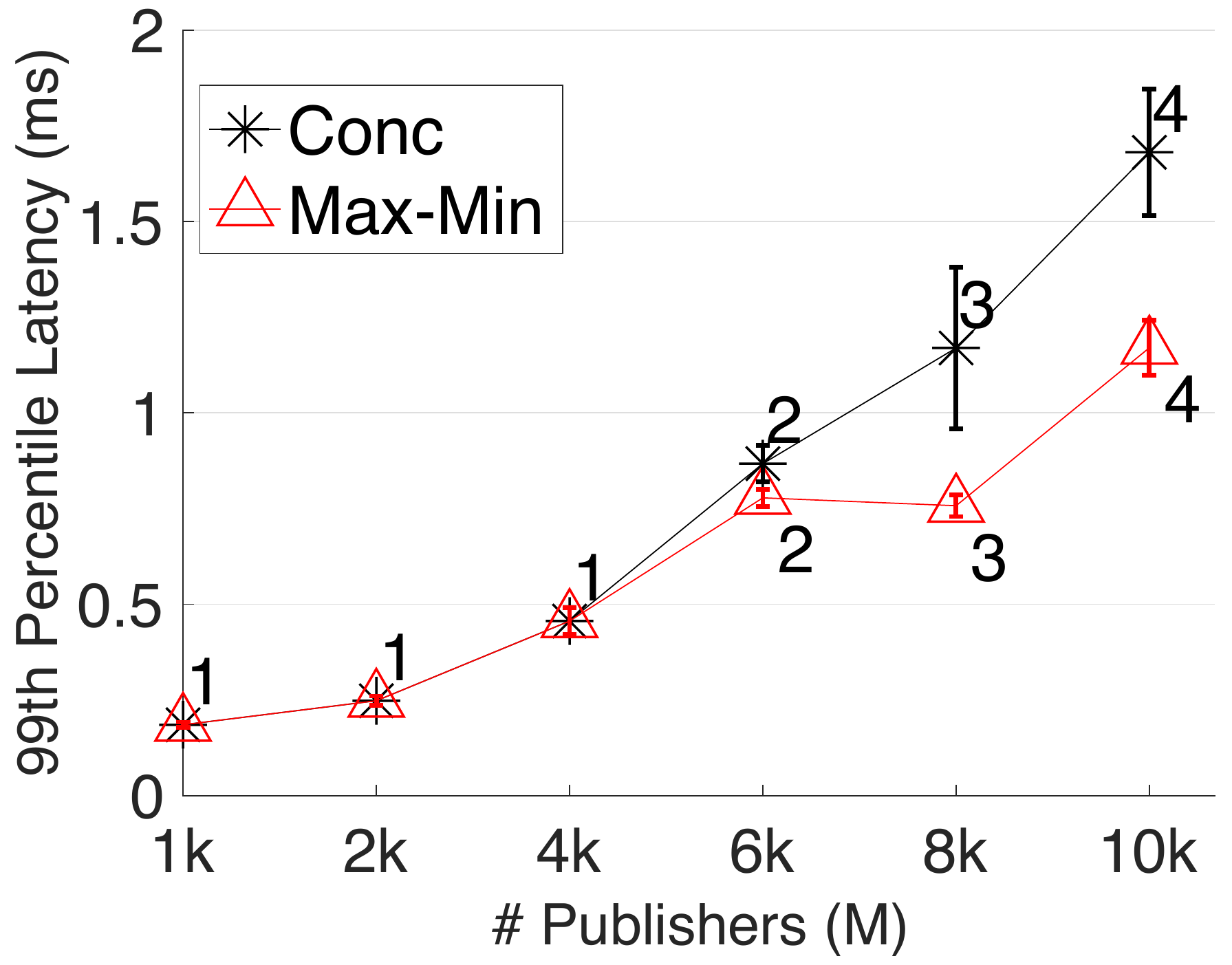}
    } 
   ~
    \subfloat[Poisson, Batch=10\label{com2-2}]{%
      \includegraphics[width=0.24\textwidth]{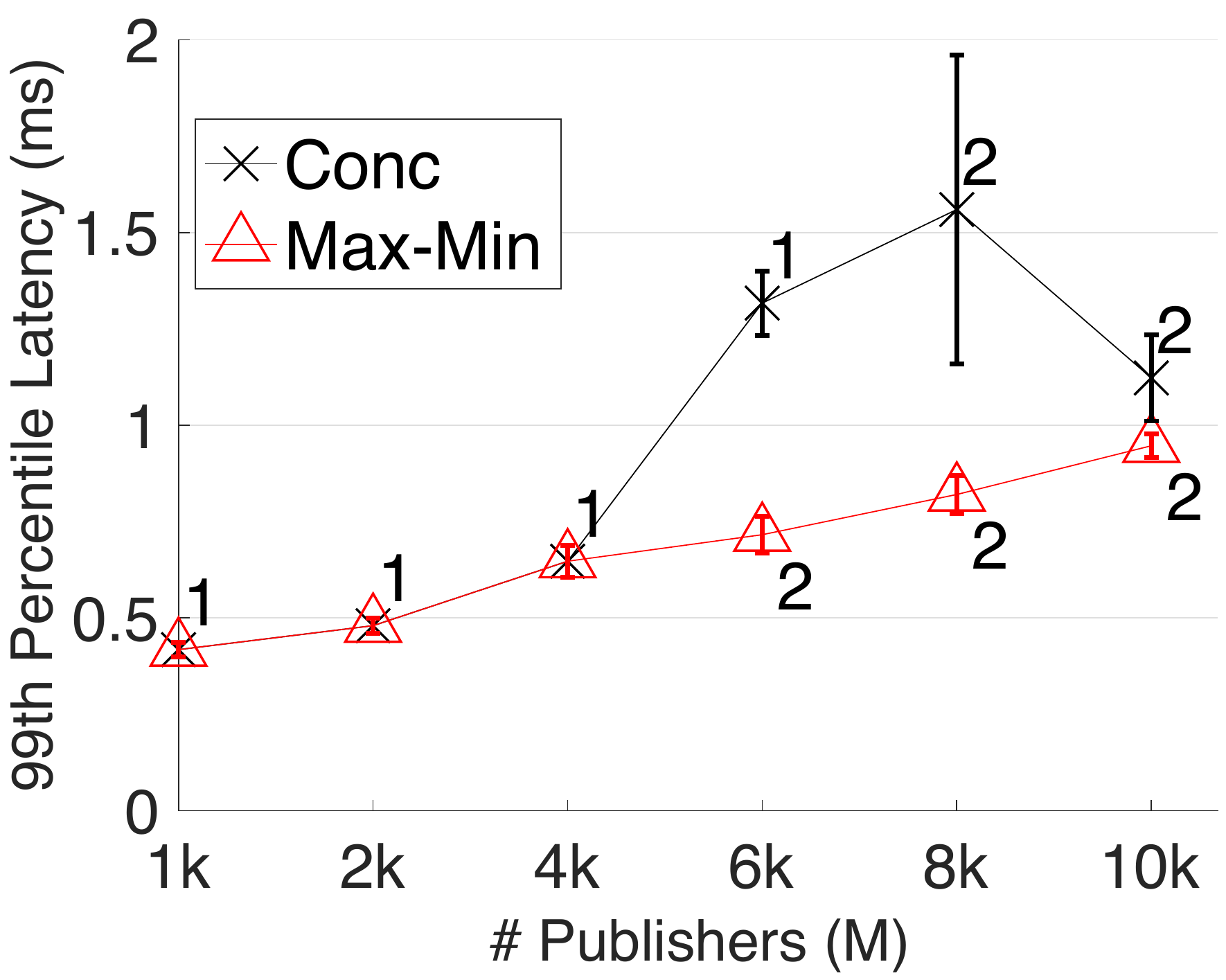}
    } 
    ~
    \newline
    ~
     \subfloat[Periodic, Batch=1\label{com2-3}]{%
      \includegraphics[width=0.24\textwidth]{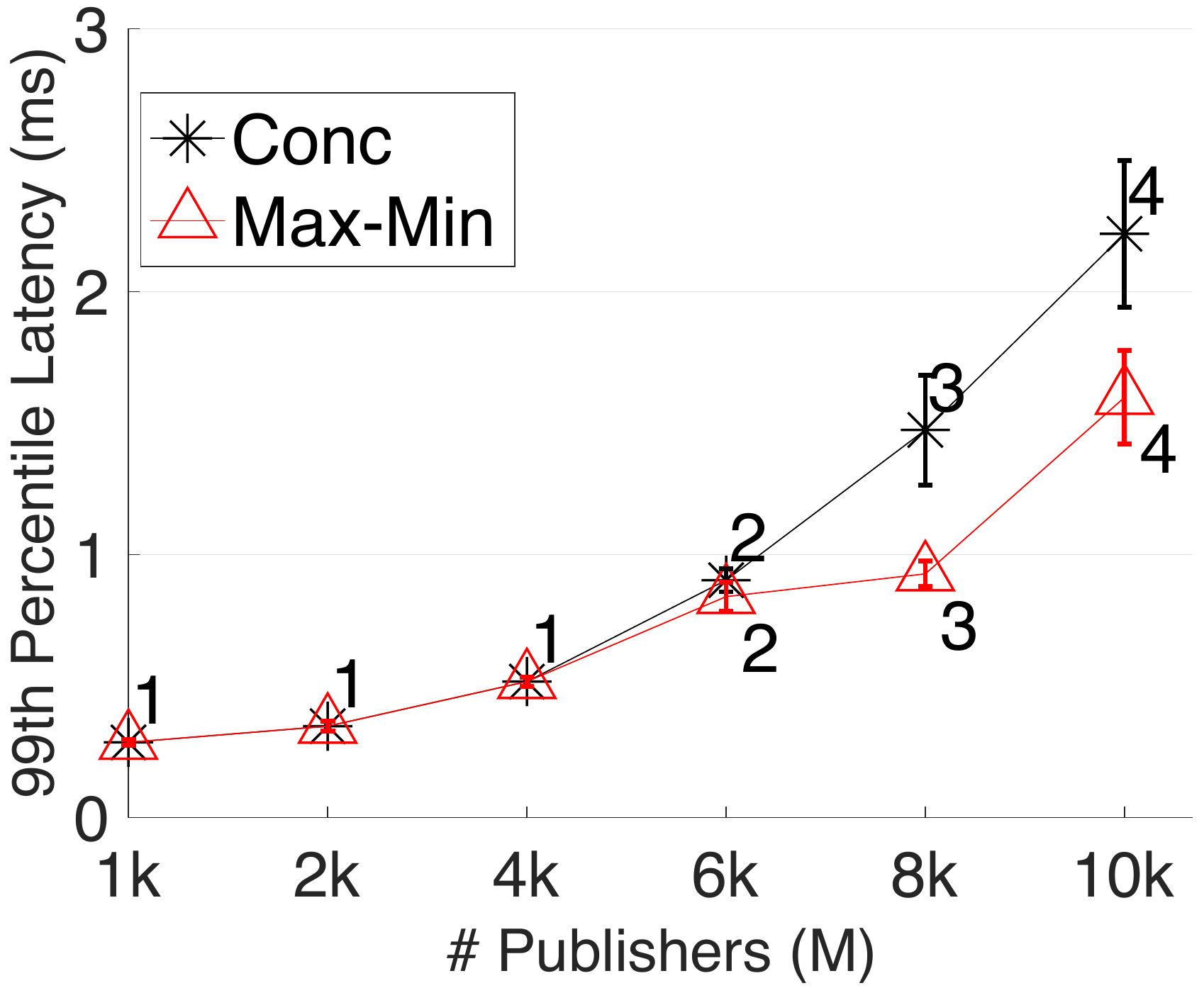}
    } 
    \subfloat[Periodic, Batch=10\label{com2-4}]{%
      \includegraphics[width=0.24\textwidth]{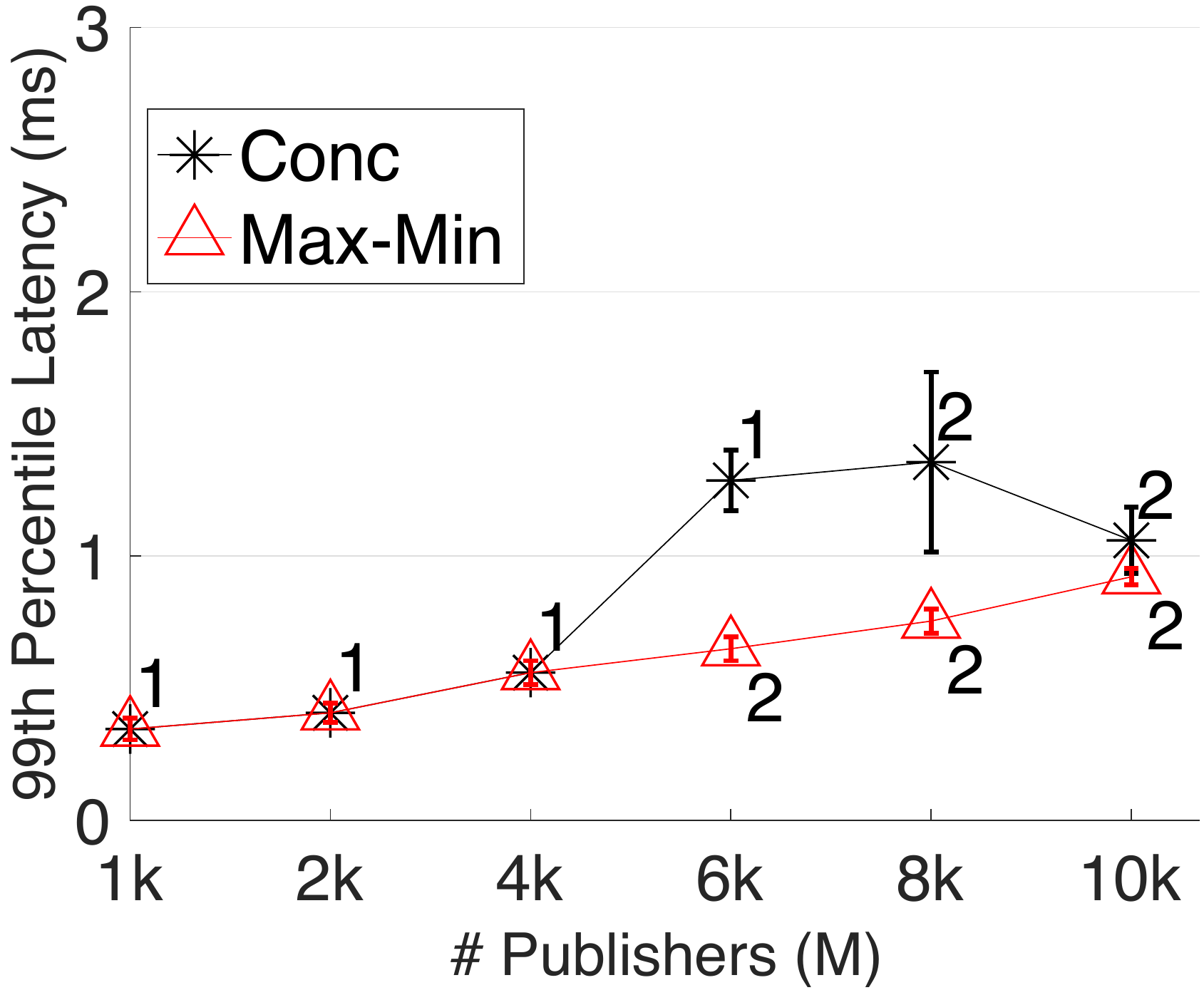}
    } 
    ~
    \caption{Impact of Max-Min on $99^{\mbox{th}}$ percentile Latency. \newline Initial (Poisson) load on brokers 1-6 (in kmsg/s): 10, 40, 50, 60, 70, 80 (w/ batch size of 10 messages). 
    }    
    \label{com2}
\end{figure}

\subsection{The Benefits of Concentration}
\label{sec: con}

In this section, we evaluate the impact of \textit{concentration} by comparing a baseline approach, \textbf{Load Balancing (LB)}, that evenly distributes a topic's publishers across all available brokers, to an alternative, \textbf{Conc}, that incorporates the \emph{concentration} principle. Conc distributes publishers across brokers so as to equalize the \emph{total} load on each broker, but unlike LB that carries this distribution out across \emph{all} brokers, it limits itself to the \emph{smallest} possible number of brokers the topic requires. This number is first estimated based on the topic's message rate and brokers' $rcap$ values, and then validated using a measurement-based approach as described in Section~\ref{sec:dist}.

The comparison is carried out for different workloads by varying the number of publishers associated with a topic.  As before, publishers have a fixed message rate of $10~msg/s$, and we again consider Poisson and periodic publishers. We vary the burstiness of the message generation process of each publisher by changing the size of the message batch they generate (1 or 10).  Periodic publishers are independent of each other, with a randomly selected phase for their period. Experiment start again with idle brokers and are repeated $10$~times.  The results are shown in \fig{com1} with the mean and standard deviation of the $99^{\mbox{th}}$ percentile latency reported for each configuration. The number of brokers across which the topic's workload is distributed is also shown next to each data point. 

\figs{com1-1}{com1-2} report the results for Poisson publishers and batch sizes of~$1$ and~$10$, respectively, with \figs{com1-3}{com1-4} devoted to periodic publishers.  Results are qualitatively consistent across scenarios, and illustrate the benefit of the \emph{concentration} principle (Conc meets the topic's SLO for all configurations, while LB consistently fails to).  The figures also highlight two relatively intuitive factors.  

The first is the impact of the aggregate message rate on the DRL penalty.  Specifically, as the number of publishers, and therefore the topic's aggregate message rate, increases, the penalty that LB incurs decreases.  A return to \Eqref{eq:tb_md1} readily explains why.  A higher token rate means a smaller token generation time (service time), and consequently a shorter delay as is well-known from basic queueing theory.  Hence, while the DRL penalty is still present, a high message rate means that the token rate at each broker even after splitting the traffic 6 ways (as LB requires) remains high enough to ensure a comparatively small penalty relative to the message processing delay. 

The other factor the figures bring to light is how traffic burstiness amplifies the DRL penalty.  This can again be explained by looking at \Eqref{eq:tb_md1}.  When applied to a batch (Poisson) arrival process, the bucket size and the token rate are scaled down by the batch size, which both contribute to an increase in delay (the last message in a batch of size~$10$ that finds an empty token bucket waits for $10$ tokens).  It is this ``amplification'' factor that is behind the significantly worse performance of LB for batch arrivals.

\begin{figure}
   \subfloat[Poisson Topic \label{com3-1}]{%
    \includegraphics[width=0.24\textwidth]{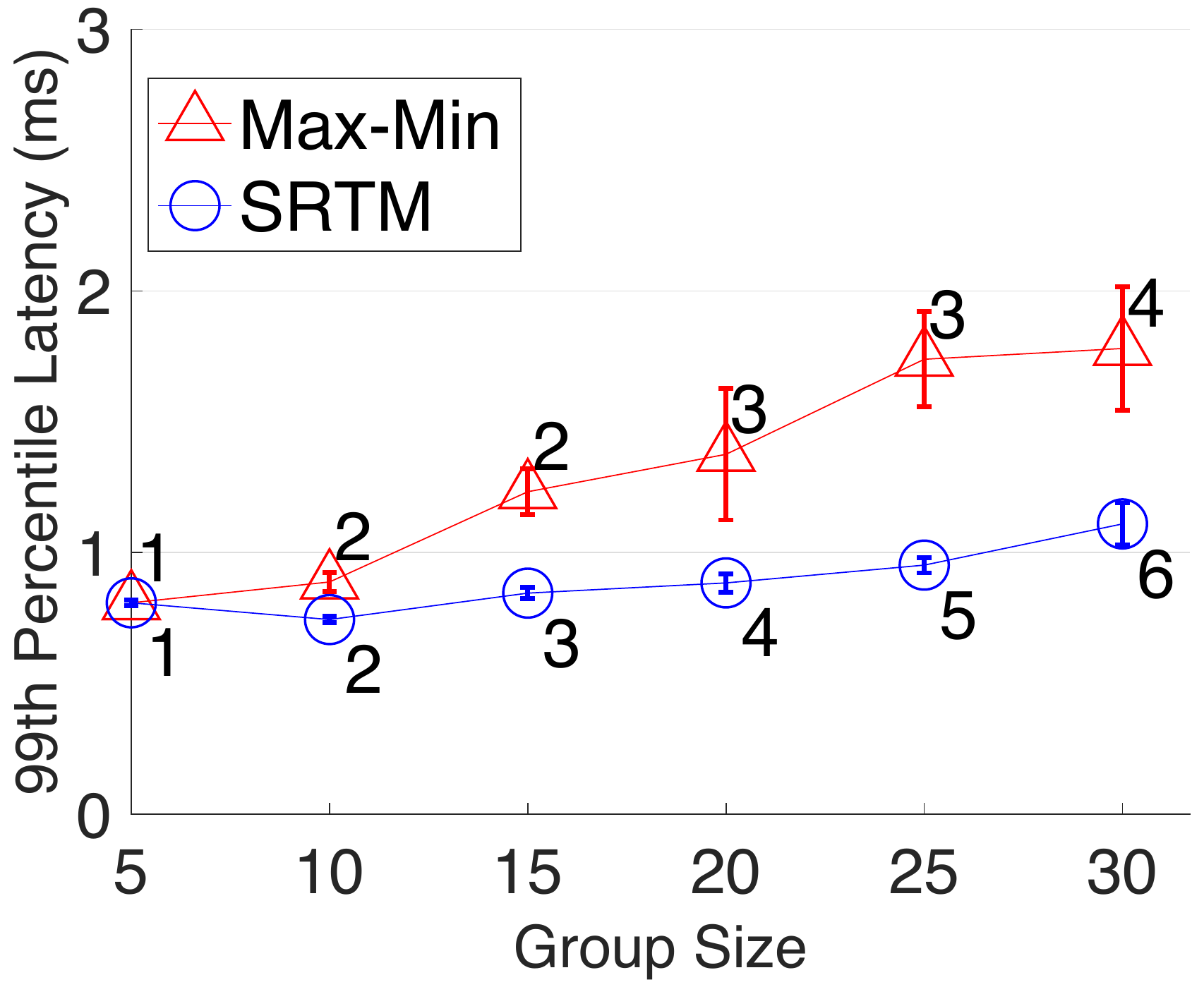}
    }
  ~
    ~
    \subfloat[Periodic Topic \label{com3-3}]{%
      \includegraphics[width=0.24\textwidth]{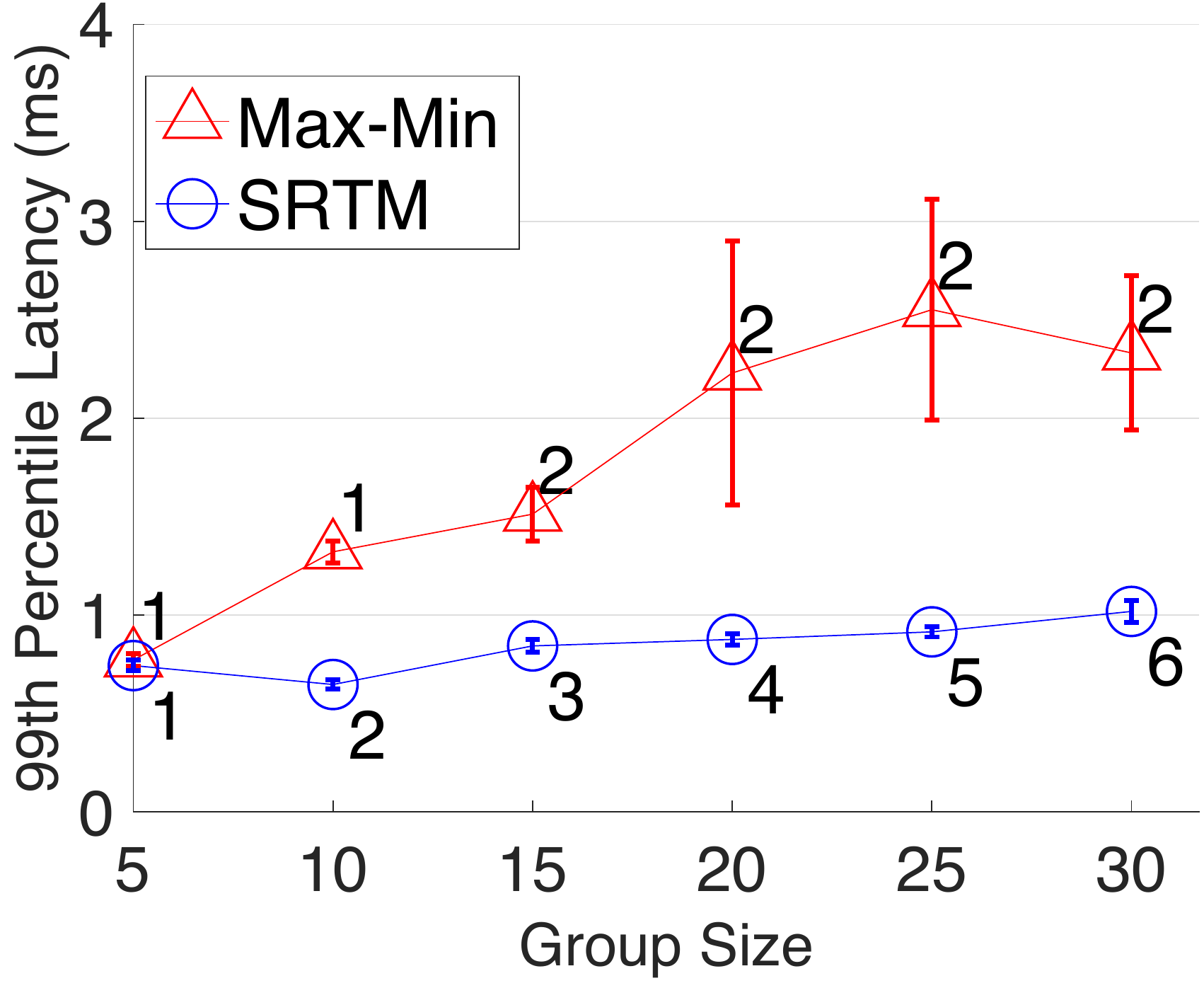}
    }
    \caption{Impact of Correlation-aware Allocation.\newline
    (Each publisher has a batch size of 10 and an average message rate of 10/s; Topic has 1,000 publishers).}
    \label{com3}
\end{figure}

\begin{figure}
   \subfloat[Poisson Topic \label{com33-1}]{%
    \includegraphics[width=0.24\textwidth]{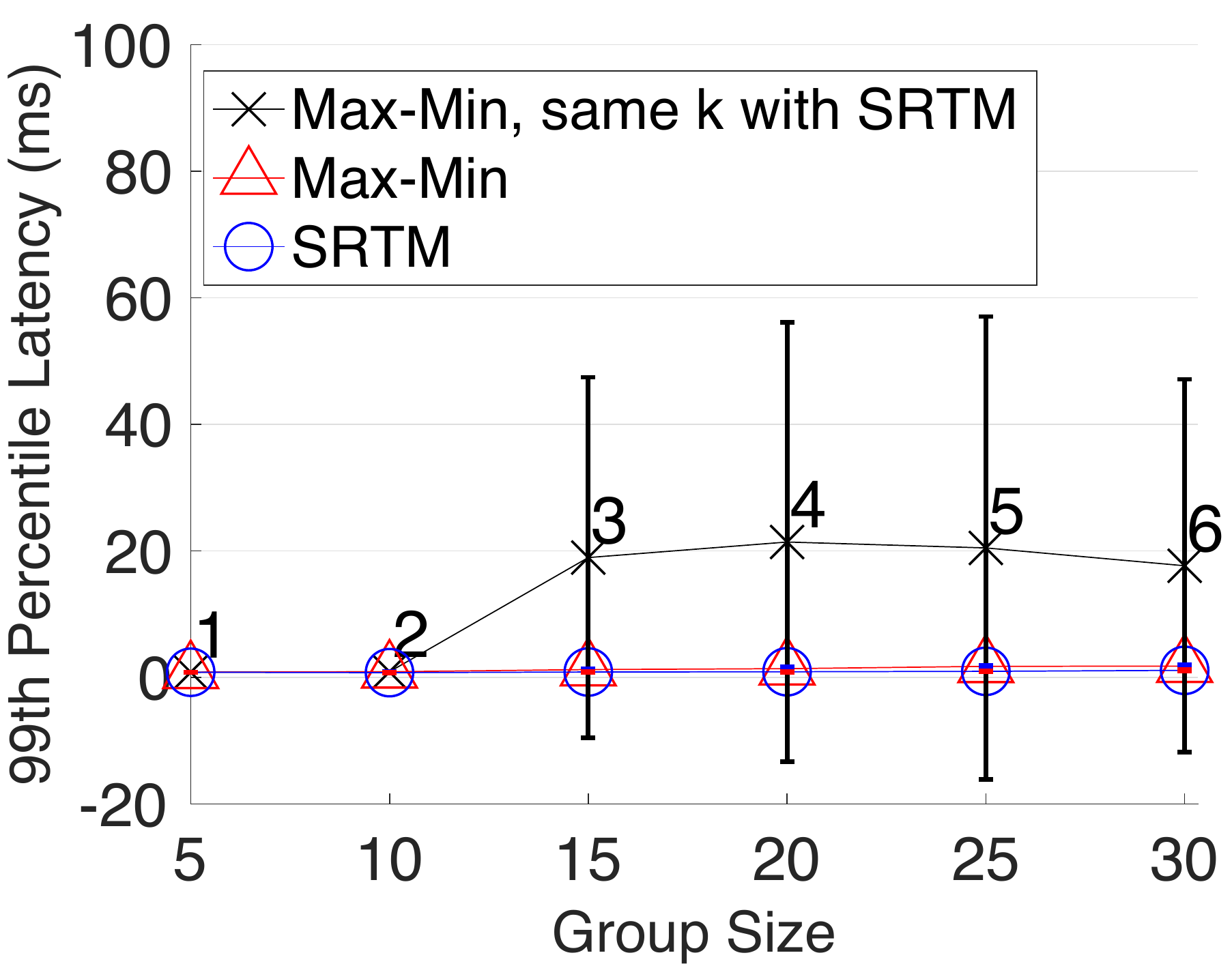}
    }
  ~
   \subfloat[Periodic Topic  \label{com33-2}]{%
      \includegraphics[width=0.24\textwidth]{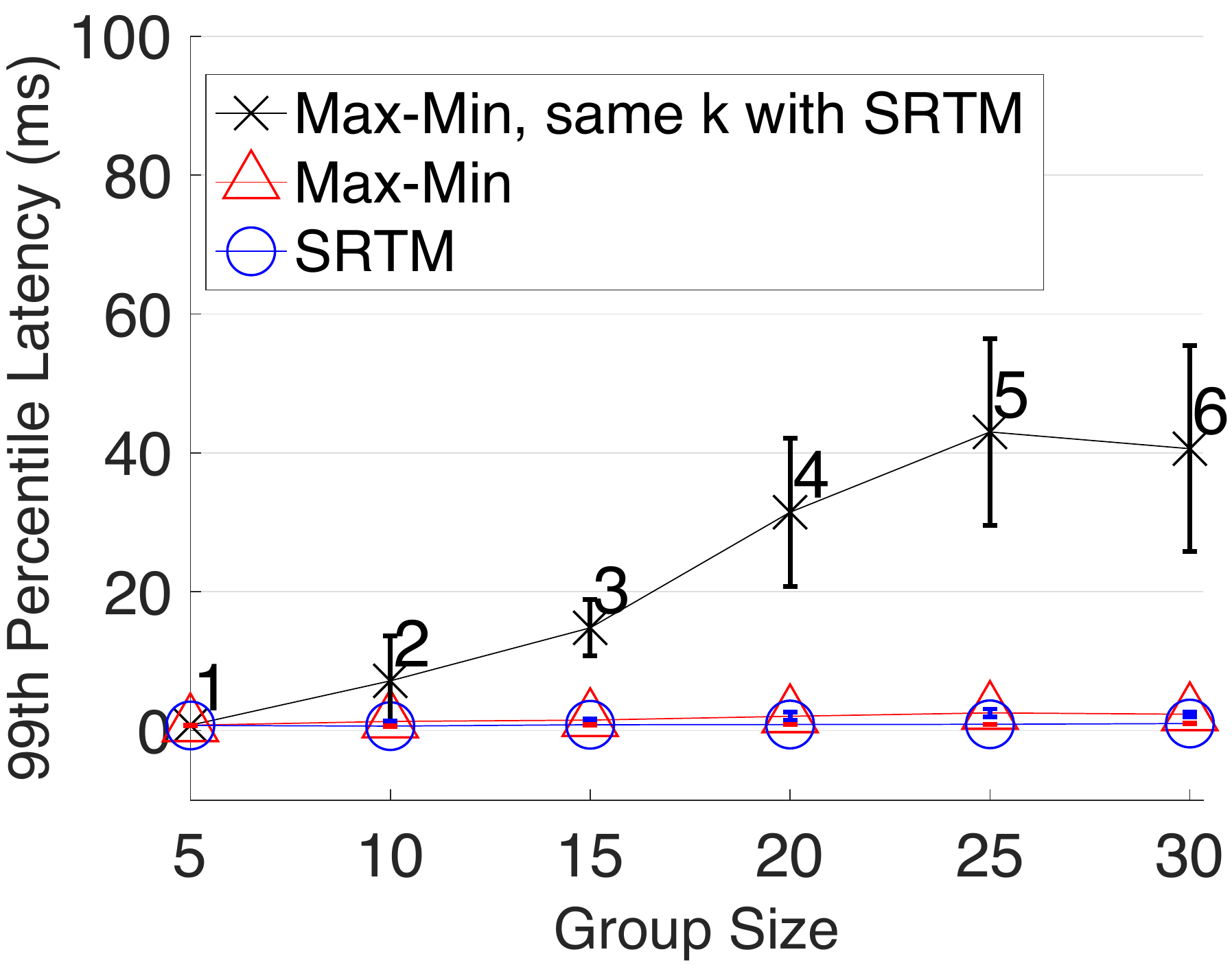}
    }
    \caption{Impact of Correlation-aware Allocation.\newline
    (Max-Min uses the same number of brokers as SRTM).}
    \label{com33}
\end{figure}

\begin{figure}
    \subfloat[Poisson\label{com4-1}]{%
     \includegraphics[width=0.24\textwidth]{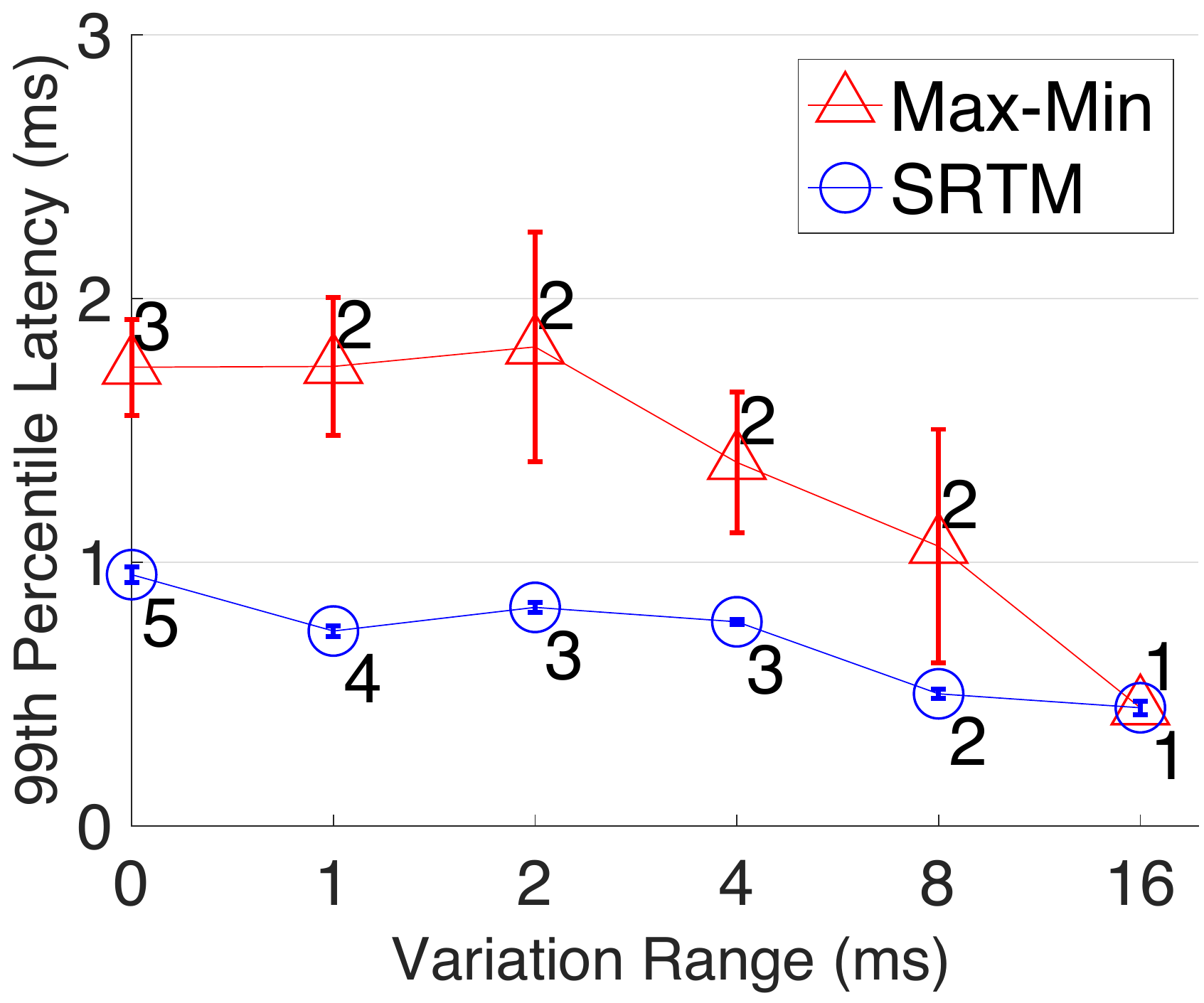}
    }
    ~
    \subfloat[Periodic\label{com4-2}]{%
    \includegraphics[width=0.24\textwidth]{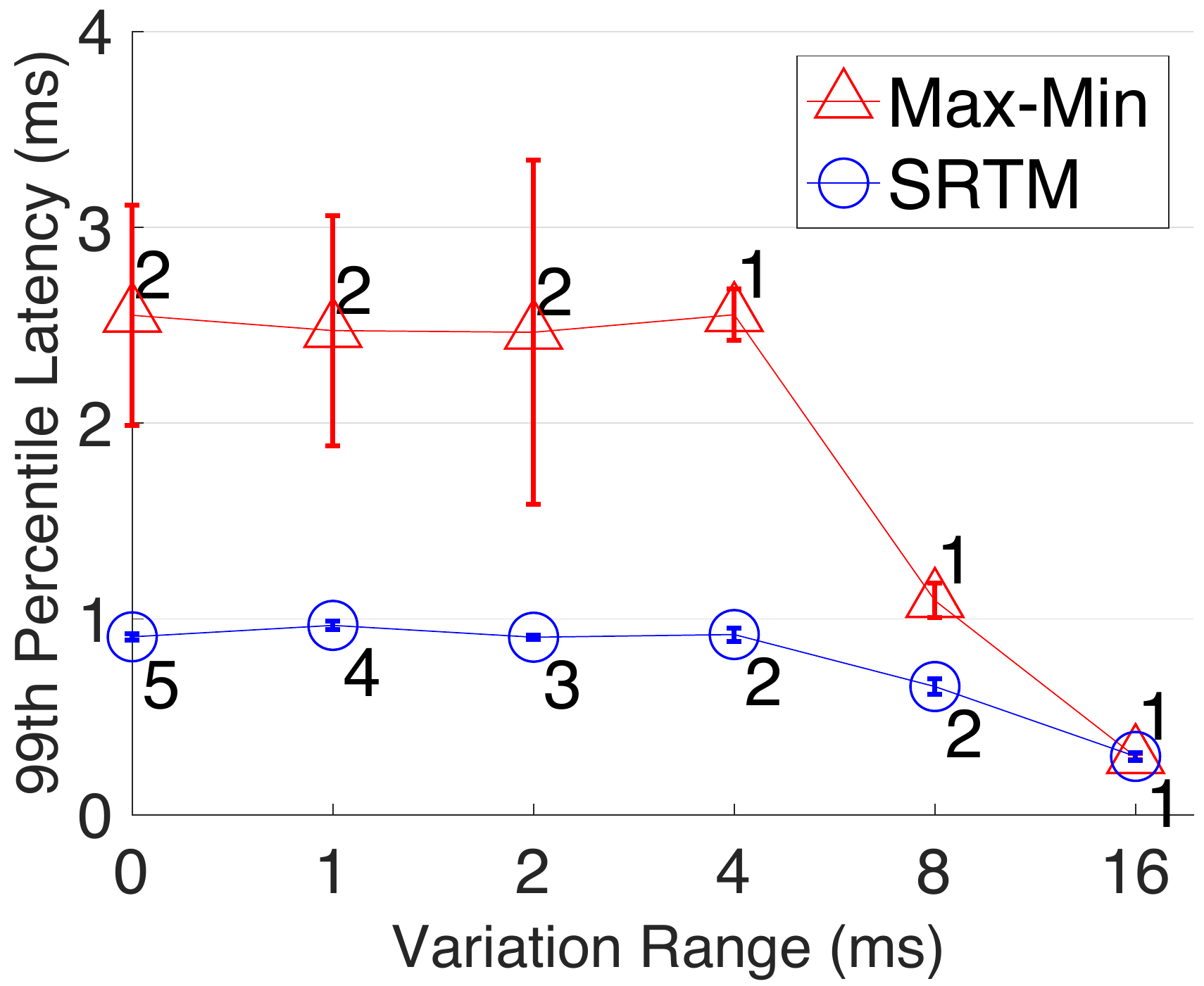}
    }
    \caption{Latency evaluation - imprecise correlation.\newline
    (Each publisher has a batch size of 10 and an average message rate of 10/s; Topic has 1,000 publishers).}
    \label{com4}
\end{figure}

\subsection{The Benefits of Max-Min}
\label{sec: max-min}

To evaluate the impact of the \textit{max-min} principle, we reuse the scenarios of the previous section, but assume that a new topic finds brokers with varying pre-existing message loads (as shown in the caption of \fig{com2}).
We also introduce a new baseline algorithm, \textbf{Max-Min}, that extends Conc by incorporating the max-min principle. As Conc, Max-Min targets accommodating the new topic with the smallest possible number of brokers, but instead of aiming to equalize the \emph{total} load at each broker, it seeks to maximize the \emph{minimum topic} load assigned to a broker.

\figs{com2-1}{com2-2} compare Conc and Max-Min for a new topic with Poisson publishers, while \figs{com2-3}{com2-4} report similar results for periodic publishers. The presence of existing workloads on the $6$~brokers affects the new topic's distribution across brokers as a function of its own workload.  When it is low, it can fit on the most lightly loaded broker and Conc and Max-Min perform identically.  Their performance, however, starts deviating as the topic's load increases (beyond 4k publishers) and needs to be split over multiple brokers.  For example, when the topic boasts 8k publishers, it now needs to be distributed across brokers 1, 2, and 3. Conc equalizes the total load of the three brokers, while Max-Min instead seeks to maximize the topic's load on broker~$3$ that, because it has the heaviest existing load, receives the smallest share.  This results in a $99^{\mbox{th}}$ percentile latency of $1.50ms$ for the publishers assigned to broker~$3$ under Conc, while it is $0.87ms$ under Max-Min, and the difference primarily arises from the larger DRL penalty under Conc.

As the two sets of figures show, Poisson and periodic publishers yield similar outcomes, but the experiments also reveal an interesting yet ultimately intuitive behavior when it comes to the impact of burstiness in the topic's arrival process.  In particular, \figs{com2-3}{com2-4} display performance for a new topic with publishers that generate bursts of $10$~messages, and both Conc and Max-Min perform as well if not better as when publishers generate bursts of size~$1$.  This initially counter-intuitive behavior is because, in this scenario, the performance of the new topic is dominated by the processing of the {\tt IOLoop} goroutines associated with its publishers (recall the overview of Section~\ref{sec: nsq}). Under a bursty arrival process, a publisher's {\tt IOLoop} is scheduled less frequently, which results in a smaller number of simultaneously active {\tt IOLoop}s that need to be serviced by the Go runtime scheduler.  This results in a broker being able to handle a higher overall message load under both Max-Min than Conc, simply from the smaller number of {\tt IOLoop}s from the new topic\footnote{As described earlier, such complex interactions in the NSQ architecture were behind the online-fitting Step~$(3)$ of Section~\ref{sec:dist}.}.

Another interesting behavior that \figs{com2-3}{com2-4} reveal is the improvement in performance of Conc when the new topic goes from 8k to 10k publishers.  The reason is again differences in DRL penalty. With 8k publishers, the ``left-over'' number of publishers assigned to the second broker is smaller than with 10k publishers, and consequently the resulting DRL penalty is higher. This illustrates the primary weakness of relying on concentration only, as it can result in residual assignments to the last broker that produce very high DRL penalties.  Again, this is the primary motivation behind the Max-Min principle.

\subsection{The Benefits of Correlation Awareness}
\label{sec:corr}

As discussed in Section~\ref{sec:correlation}, correlation in how publishers generate messages is captured through correlation groups.
We then compare Max-Min to \textbf{SRTM} that incorporates group information when distributing publishers across brokers (Max-Min is oblivious to that information, while SRTM seeks to leverage it to distribute publishers across brokers so as to reduce arrival burstiness in proportion to the reduction in bucket size).  Towards isolating the impact of correlation, we again assume that a new topic arrives to a set of idle brokers.

Additionally, because correlation groupings can be coarse, \eg reflecting physical proximity rather than precise synchronization, we consider two scenarios.  The first assumes that publishers within the same group are perfectly correlated, \ie their message generation times are precisely synchronized, while the second relaxes this assumption by introducing variations in the times at which publishers in the same group generate messages. 

\subsubsection{Scenario~$1$: Accurate correlation}
\label{sec: precise}
In this set of experiments, publishers marked as belonging to the same group are perfectly synchronized in their message generation times.  We vary group sizes across experiments, with larger groups corresponding to larger (synchronized) message bursts generated by each group.  The main consequence of such an increase is that meeting the $1ms$ SLO for burstier traffic calls for distributing the topic across more brokers.  This affects both Max-Min and SRTM, but the fact that SRTM relies on group information to split publishers from the same group across brokers enables it to mitigate the resulting increase in DRL penalty.  

The results are shown in \figs{com3-1}{com3-3} for Poisson and periodic publishers, respectively.  Both figures illustrate that SRTM is able to gain access to more broker capacity (and break message bursts) without incurring a significant increase in DRL penalty.  This allows it to meet the target SLO, even for groups of size~$30$.  In contrast, Max-Min is forced to use fewer brokers, as its ``blind'' assignment of publishers ultimately results in a DRL penalty that exceeds the benefits of distributing the topic's message load across more brokers.  This is further illustrated in \fig{com33} that also reports the performance of Max-Min when it tries to use the same number of brokers as SRTM.  As anticipated, this makes its performance even worse.

\subsubsection{Scenario~$2$: Noisy correlation}

This next set of experiment is explores the extent to which the benefits of a correlation-aware distribution of publishers remain when correlation information is inaccurate.  Specifically, the message generation times of publishers within the same group are now (evenly) spread over an interval instead of perfectly concurrent. As the interval size increases, correlation between publishers weakens.

Results are reported in \fig{com4} that compares SRTM and Max-Min, again for both Poisson and periodic publishers.  The experiments relied on the same type of publishers as in the previous section with a group size set to~$25$, and the interval across which messages from the same group were distributed ranged from $0ms$ to $16ms$ ($0ms$ corresponds to the configuration of the previous section). The results are again consistent for both Poisson and periodic publishers and demonstrate that, at least when the level of noise is small (of the order of a few percent of the average message inter-arrival time), leveraging correlation information still helps mitigate the DRL penalty.  The figure also illustrates another side-effect of increasing the size of the interval over which publishers' messages arrive, namely tail latency as well as the number of brokers across which publishers are distributed decrease for both SRTM and Max-Min.  This behavior is a direct consequence of the lower burstiness associated with the increased ``spreading'' of message arrivals.

\subsection{Load Distribution Latency}
\label{sec:time}

As described in Section~\ref{sec:dist}, SRTM uses only the broker's average message load to estimate the residual capacity $rcap$ available on each broker. This simple approach is because the complex internal architecture of NSQ makes accurately modeling the impact of higher order arrival statistics challenging if not impossible.  As a result, SRTM resorts to a measurement-based solution to ``fine-tune'' its initial (imprecise) capacity estimates and the resulting publishers to brokers allocations. This fine-tuning is carried out in the online-fitting component of Step~$(3)$ of the load distribution mechanism.  As previously mentioned, a potential disadvantage is that such an approach can take time to converge.  Gaining insight into this \textit{load distribution latency} is the purpose of this section. 

Before reporting the results of those experiments, we however highlight that significant improvements are possible if topics' workloads are limited to a few well-understood traffic profiles for which customized models can be developed (as may be the case for specific applications).  Such specialization is, however, beyond the scope of this paper, and instead we proceed next to quantify the load distribution latency in one illustrative configuration.

Specifically, we consider scenarios involving a new topic arriving to a set of empty brokers.  This latter assumption arguably simplifies the load distribution process, as the brokers are homogeneous in their spare capacity.  Nevertheless, the basic steps of SRTM's load distribution remain present.  The topic's message load stands at a relatively low level of $10k~msg/s$ ($1,000$ publishers, each with a message rate of $10~msg/s$ and a burst size of 10), and we vary its burstiness by changing the publishers' group size (publishers in a group are perfectly synchronized).  When burstiness is low (small group size), a single broker is able to accommodate the new topic, but as burstiness increases, so does the number of brokers needed, with $6$~brokers eventually required when the group size reaches $30$.  

Because the topic's workload is low and brokers are initially empty, the first $rcap$-based assignment always starts with $k=1$ (a single broker is deemed to have enough capacity).  Depending on the topic's burstiness, additional brokers may, however, be needed (resulting in an increase in $k$). This is reflected in the different rows of Table~\ref{step}, where each row corresponds to a different level of burstiness, and within a given row, a column ($k$ value) represents one iteration of Step~$(3)$ of the load distribution. 

An iteration starts with an assignment of publishers to a broker\footnote{In the first iteration, all publishers are assigned to broker~$1$.} and seeks to assess if the SLO is met.  If it is, the load distribution process successfully completes.  If it is not, a binary search is initiated to determine the maximum number of publishers the broker can accommodate\footnote{In our experiments, the search stops as soon as latency is within $20\%$ below the SLO target of $1ms$.}.  Entries in the table give the number of search rounds. Each round lasts $60$~secs to ensure a representative traffic sample and are the primary contributors to load distribution latency, whose total value is reported in the last column.  Once the number of publishers a broker can accommodate has been identified, excess publishers are then assigned to the next broker and $k$ is increased by~$1$.  

We note that because brokers are initially idle, iterations always proceed one broker at the time, with each broker fully loaded before the next one is added, \ie we never have to consider shifting excess publishers to previously considered brokers.  This is also why the last iteration completes in a single round.

Table~\ref{step} indicates that as the burstiness (group size) of the topic increases, so does the duration of the load distribution phase.  This is expected, since our estimate of $rcap$ is oblivious to workload burstiness, and therefore becomes less accurate as it increases.  For bursty topics, initial assignments systematically over-estimate the number of publishers a broker can accommodate. This triggers repeated iterations, each calling for a binary search. In the ``worst'' case (group size of 30), the cumulative effect of those searches results in a load distribution phase in excess of $15$~minutes.  This is clearly long, though not unreasonable when dealing with, say, an IoT deployment that may last for weeks or months.  Additionally and as mentioned earlier, if variations in traffic profile parameters can be constrained, better models for estimating $rcap$ are feasible and would reduce the number of iterations. 



\begin{table}[t]
\begin{center}
    \begin{tabular}{ | p{0.6cm} | p{0.5cm} | p{0.5cm} | p{0.5cm} | p{0.5cm} | p{0.5cm} | p{0.5cm} | }
    \hline
    \footnotesize{\textbf{Group Size}} & \footnotesize{$k$\newline$=1$} & \footnotesize{$k$\newline$=2$} & \footnotesize{$k$\newline$=3$} & \footnotesize{$k$\newline$=4$} & \footnotesize{$k$\newline$=5$} & \footnotesize{$k$\newline$=6$} \\ \hline    
    5 & 1 & & & & & \\ \hline 
    10 & 2& 1& & & & \\ \hline 
    15 & 5 &3 &1 & & & \\ \hline 
    20 & 5 &4 &2 &1 & & \\ \hline
    
    25 & 3 &4 &4 &2 &1 & \\ \hline
    30 & 3 & 5 &3 &2 &3 &1 \\ \hline
    \end{tabular}
    \hspace{0.001cm}
    \begin{tabular}{ | p{0.5cm} |}
    \hline
    \footnotesize{\textbf{Time (sec)}}\\ \hline    
    64 \\ \hline 
    189 \\ \hline 
    558 \\ \hline 
    745 \\ \hline
    
    871 \\ \hline
    1058 \\ \hline
    \end{tabular}

\end{center}
\caption{Load distribution latency of SRTM \newline $k$: iteration index; each $k$ column gives the number of measurement rounds in that iteration; the last column reports the total load distribution time.}
\label{step}
\end{table}

\section{Related Work}
\label{sec:related}

On the system side, distributed rate limiting (DRL) has been realized in a number of cloud services through software platforms such as Cloud Bouncer~\cite{yahoo}, Tyk~\cite{tyk} and Doorman~\cite{doorman}. Those platforms provide coordination protocols that set rate limiting parameters across resources as a function of a global contract and workload partitions. 

Also relevant to our work are approaches that dynamically \emph{adapt} control decisions across distributed resources based on a global objective.
Raghavan et al.~\cite{barath} describe a system of (distributed) rate limiters that drop packets probabilistically to emulate the effect of a single aggregate rate limiter on transport layer congestion. 
Retro~\cite{retro} proposes a resource management framework that can enforce objectives such as fairness or latency guarantees by dynamically adapting resource allocation across multiple systems and tenants. It relies on a centralized controller coupled to a measurement infrastructure and distributed control points.
Stanojevic et al.~\cite{stanojevic09a, stanojevic09b} develop analytical models that account for both load balancing and rate limiting in determining how to optimally control access to distributed resources as a function of demand. The goal is to realize a fair outcome while incurring a low computational and communication overhead. 

As alluded to in Section~\ref{sec:background}, the primary difference between those works and the approach we adopt is that they seek to replicate the behavior of a centralized system in spite of their distributed nature, while we instead acknowledge the impact of distribution (the DRL penalty) and attempt to mitigate it.  Either approach has its own merit with benefits that vary as a function of the environment in which it is deployed.  We believe that the mostly static partitioning of resources on which we rely\footnote{The time-scale of adaptation of the TB adaptor is coarse.} affords robustness in settings where adaptation latency is large compared to the time-scale of workload dynamics.


Another related body of work targets the same \emph{concentration} principle as SRTM, but to reduce operational or energy costs while meeting SLOs.
WorkloadCompactor~\cite{wcompactor} explores how to best shape (through token buckets) co-located workloads with latency constraints to maximize the number of workloads that a server can accommodate. A similar cost reduction goal is behind the use of concentration in STeP~\cite{step} that realizes it by co-locating tenants with compatible resource usage patterns. Other systems~\cite{borg, bistro, central, scavenger} realize a similar goal by co-locating latency-sensitive workloads with data-intensive (batch-processing) workloads. 
In spite of similarity in the end-goal (concentration), our goal of seeking it to mitigate the DRL penalty differs from that of those earlier works.

Similarly, load-balancing strategies are common in distributed cloud services. The objective is to evenly distribute an \textit{aggregate} load across servers to optimize performance given an available set of servers.  Load-balancers~\cite{maglev,cheetah,pesto,romano} can therefore be effective in minimizing processing latency.  However, as illustrated in \fig{fig:2configs} and demonstrated in \fig{single}, a strict load-balancing strategy can be counter-productive in the presence of a DRL penalty.  This is true even when combining it with concentration, which motivated the introduction of the \emph{Max-Min} principle in SRTM.


Finally, leveraging workload \emph{correlation} to improve system performance has been the focus of a number of earlier works, which have typically relied on traces to
capture correlation in resource usage. For example, STeP~\cite{step} used FFT co-variance to reduce CPU contention by co-locating anti-correlated databases.  Power consumption was the target of Verma et al.~\cite{server} that proposed to consolidate anti-correlated applications based on Pearson correlation, and of Carpo~\cite{carpo} that also relied on Pearson correlation to consolidate loosely-correlated flows in data center networks. Although SRTM shares with those works a reliance on correlation to optimize performance, it differs in both the performance metric it targets (the DRL penalty) and its approach to measuring correlation (using IoT application semantics rather than measurements).





\section{Conclusion}
\label{sec:concl}
The paper's main contributions are in identifying and explicating the DRL penalty that unavoidably arises when rate control is distributed across servers,
and in designing and developing a system, SRTM, capable of mitigating this penalty with a focus on its use by IoT applications.  

The SRTM system relies on three core principles: \textit{concentration}, \textit{max-min} and \textit{correlation-awareness}, and was evaluated empirically on a local testbed. The evaluation demonstrated its ability to successfully mitigate the DRL penalty, while preserving the ability to scale by distributing workload across servers when needed. SRTM was developed on top of the NSQ open-source messaging platform and is publicly available for others to use~\cite{srtm}.

SRTM is fully operational, but relies on a measurement-based solution to accurately match arbitrary workload to system resources.  This matching can in some cases be time-consuming.  A promising extension for configurations involving more specialized workloads involves developing an accurate model-based resource matching solution that would overcome this limitation.

\clearpage
\bibliographystyle{abbrv}
\bibliography{ref}

\clearpage
\appendix
\section{Rate Limiting Delay}
\label{app:2vs1}
\subsection{Model and Assumptions}

Consider a two-parameters token bucket~\cite{rfc2697} $(r,b)$ where
$r$ denotes the token rate (in messages/sec) and $b$ the allowed burst
size (in messages). In other words the number of messages that can
leave the token bucket in any time interval of duration $\Delta t$
(the arrival curve to the system downstream of the token bucket) is
upper-bounded by $b+r\cdot \Delta t$.  Message arrivals to the token
bucket follow an arbitrary arrival process and each message consumes
one token.  Messages that find an available token upon their arrival
immediately clear the token bucket without incurring any
delay. Messages that arrive to an empty token bucket (or a token
bucket with only a fraction of a token) wait until a full token is
available before they are allowed to leave the token bucket. The
waiting space at the token bucket is assumed large enough (infinite)
to ensure that messages waiting for tokens are never lost.

Of concern are delays incurred in the token bucket, \ie
the rate limiting delays.  Our focus is on the sum\footnote{Note that
  the sum of message delays divided by the number of messages gives
  the ensemble mean message latency (up to time~$t$).} $S(t)$ of the
delays accrued by all message up to time $t$.  In particular, we are
interested in the impact on $S(t)$ of replacing a one-bucket system
$(r,b)$ with a two (or more) bucket system consisting of two separate
sub-token buckets $(r_1,b_1)$ and $(r_2,b_2)$, where $r=r_1+r_2$ and
$b=b_1+b_2$.  In the two-bucket system, the original stream of
arrivals is split arbitrarily across the two sub-token buckets at the
times of message arrivals, and each sub-token bucket has an infinite
queue where messages waiting for tokens can be stored. 

We note that under the assumptions of a general message arrival
process with each message requiring exactly one token, a token bucket
system with unit token rate, \ie $r=1$, and a bucket size of $b$
tokens behaves like a modified G/D/1 queue with unit service times.
The modification is that in the token bucket system, messages
experience a delay if and only if the queue content in the G/D/1
system exceeds $b-1$.  In other words, assuming a
first-served-first-come (FCFS) order of service\footnote{Note that a
  FCFS service order is known to minimize the sum of message latencies
  in both single-server and (homogeneous) multi-server systems when
  service times are constant~\cite{uuganbaatar11}.}, the token bucket
delay $d_i$ of the $i^{th}$ message can be obtained from the system
time of this message in the corresponding G/D/1 system as follows:
\beq
\label{eq:tb_delay}
d(a_i)=\max\{0,U(a_i^-)+1-b\}\, ,
\eeq
where $U(a_i^-)+1$ corresponds to the unfinished work found in the
G/D/1 queue by the $i^{th}$ message upon its arrival at time $a_i$
plus its own contribution to the unfinished work, and $b$ is the
bucket size.

Next, we proceed to compare the relative (latency) performance of a
one-bucket system to that of a multi-bucket system obtained by splitting
the one-bucket system as described above.  In particular, we establish
that splitting a token bucket in two (or more) sub-token buckets
always increases the sum of the message delays, and hence also the
average message delay.

\subsection{One vs.~Two or more Token Buckets}
\label{app:split}

Towards establishing the result that splitting a token bucket can only
worsen the sum of message delays, we first state a simple Lemma.
\begin{lemma}
\label{lemma:uw}
At any point in time $t$, the unfinished work $U(t)$ in a
work-conserving G/D/1 queue is smaller than or equal to the total
unfinished work $U^{(k)}(t)=\sum_{l=1}^kU_l(t)$ in a set of $k$
work-conserving G/D/1 queues with the same aggregate service rate and
fed the same arrival process.
\end{lemma}
\begin{proof}
The result directly stems from the observation that when fed the same
arrival process, $k$ parallel work-conserving G/D/1 queues never clear
work faster than a single work-conserving G/D/1 queue with the same
aggregate service rate.  Specifically, at any point in time both the
one-queue and the $k$-queues system have received the same amount of
work (they are fed the same set of arrivals), both systems are
work-conserving, and the one-queue system processes work at least as
fast as the $k$-queue system whenever it is not empty, so that it can
never have more unfinished work than the $k$-queue system.

Formally, we assume that up to the start of the $j^{th}$ busy period
of the one-queue system, the unfinished work in the one-queue system
has always been smaller than or equal to that of the $k$-queue
system, and wlog we assume that the one-queue system has unit service
rate.  We establish the result by induction on the busy periods of the
one-queue system.

Denote as $t_j$ the start of the $j^{th}$ busy period of the one-queue
system, and let $T_j$ denote the duration of that busy period.  The
unfinished work in the one-queue system during that busy period is
then of the form $U(t)=U(t_j^-)+W(t_j,t)-(t-t_j)=W(t_j,t)-(t-t_j)\,
,\,\forall t\in[t_j,t_j+T_j]$, where $W(t_j,t)$ represents the amount
of work that has arrived in $[t_j,t]$, and we have used the fact that
by definition the unfinished work just before the start of a busy
period is~$0$.  Similarly, the unfinished work in the $k$-queue system
is of the form $U^{(k)}(t) = U^{(k)}(t_j^-)+W(t_j,t)-
\int_{t_j}^tr^{(k)}(u)du\geq W(t_j,t)-(t-t_j)=U(t)\, ,\,\forall
t\in[t_j,t_j+T_j]$, where we have used the facts that
$U^{(k)}(t_j^-)\geq U(t_j^-)=0$ (from our induction hypothesis),
$r^{(k)}(u)\leq 1$, \ie the aggregate service rate in the $k$-queue
system can never exceed the unit service rate of the one-queue system,
and both systems receive the same amount of work $W(t_j,t)$.
Furthermore, because by definition of a busy period $U(t_j+T_j)=0$ and
both the one-queue and the $k$-queue system see the same arrivals, we
also have $0=U(t)\leq U^{(k)}(t)\, ,\,\forall t\in[t_j+T_j,t_{j+1})$,
  where $t_{j+1}$ is the start time of the $(j+1)^{th}$ busy period of
  the one-queue system, \ie the time of the next arrival after
  $t_j+T_j$.  This establishes that the unfinished work in the
  one-queue system remains smaller than or equal to that in the
  $k$-queue system until the start of the $(j+1)^{th}$ busy period of
  the one-queue system.  This completes the proof of the induction
  step.
\end{proof}
We are now ready to prove Proposition~\ref{theo:1vs2}, which we
restate next for completeness.  

\noindent
{\bf Proposition~\ref{theo:1vs2}}
\emph{Given a two-parameter token bucket $(r,b)$ and a general message
arrival process where messages each require one token to exit the
bucket, splitting this one-bucket system into multiple, say, $k$,
sub-token buckets with parameters $(r_l,b_l)$ such that
$r=\sum_{l=1}^kr_l$ and $b=\sum_{l=1}^kb_l$, can never improve the
running sum of the message delays, irrespective of how messages are
distributed to the $k$ sub-token buckets.  More generally, denoting as
$S(t)$ and $S^{(k)}(t)$ the sum of the delays accrued by all messages
up to time $t$ in the one-bucket and $k$-bucket systems, respectively,
we have}
\beq
S(t)\leq S^{(k)}(t)\,, \,\forall t \tag{\ref{eq:theo}}
\eeq
%
\begin{proof}
We first establish the result for the case $k=2$, and wlog assume that
$r=1$. 

The proof is simply based on the fact that messages waiting for tokens
in either system accrue delay at the same rate, and establishing that
at any time $t$ the number $N(t)$ of messages experiencing delays in
the one-bucket system is less than or equal to the number
$N_1(t)+N_2(t)$ of such messages in the two-bucket system. Note that
the sum of the message delays incurred in either system up to time $t$
is of the form
\bearun
S(t) &=& \int_0^tN(u)du \\
S^{(2)}(t) &=& \int_0^t\left(N_1(u)+N_2(u)\right) du
\eearun
Hence, if $N(t)\leq N_1(t)+N_2(t)\,, \,\forall t$, then $S(t)\leq
S^{(2)}(t)\,, \,\forall t$, which proves the result for $k=2$.  We
therefore proceed to establish that $N(t)\leq N_1(t)+N_2(t)\,,
\,\forall t$.

The number $N(t)$ of messages waiting for tokens, \ie accruing delay,
at time $t$ in a one-bucket system with bucket size $b$ is of the
the form 
\bequn
N(t)=\left\lceil \max\{0,U(t)-b\}\right\rceil \, ,
\eequn
where $\lceil x\rceil$ represents the ceiling of~$x$, $U(t)$ is the
unfinished work in the corresponding G/D/1/ queue, and consistent with
\Eqref{eq:tb_delay} we have used the fact that messages are delayed in
the token bucket only when the unfinished work in the G/D/1 queue
exceeds the bucket size~$b$.

Similarly the total number of messages waiting for tokens in a
two-bucket system with bucket sizes $b_1$ and $b_2$ such that
$b=b_1+b_2$ is of the form
\bequn
N_1(t)+N_2(t)=\left\lceil \max\{0,U_1(t)-b_1\}\right\rceil
             +\left\lceil \max\{0,U_2(t)-b_2\}\right\rceil
\eequn
Since we know that $\lceil x\rceil\leq \lceil x_1\rceil+\lceil
x_2\rceil$, when $x\leq x_1+x_2$, we focus on establishing that 
\beq
\label{eq:max}
\max\{0,U(t)-b\}\leq \max\{0,U_1(t)-b_1\}+\max\{0,U_2(t)-b_2\}
\eeq
From Lemma~\ref{lemma:uw}, we know that $U(t)\leq U_1(t)+U_2(t)$.
Next, we consider the cases $U(t)-b\leq 0$ and $U(t)-b>0$.

\noindent
{\bf Case~$1$}: $U(t)-b\leq 0$

In this case, \Eqref{eq:max} is trivially verified.

\noindent
{\bf Case~$2$}: $U(t)-b>0$

We further separate this case in two separate sub-cases:

\noindent
{\bf Case~$2a$}: $U_1(t)-b_1\leq 0$ and $U_2(t)-b_2\geq 0$ (or interchangeably
$U_1(t)-b_1\geq 0$ and $U_2(t)-b_2\leq 0$) 

In this case, \Eqref{eq:max} simplifies to
\bequn
U(t)-b\leq U_2(t)-b_2
\eequn
Applying again the result of Lemma~\ref{lemma:uw}, we have
\begin{align*}
U(t)\leq U_1(t)+U_2(t)\,&\Rightarrow \, U(t)-b\leq
U_1(t)+U_2(t)-b_1-b_2\\
&\Rightarrow \, U(t)-b\leq U_2(t)-b_2\, ,
\end{align*}
where we have used the fact that $b=b_1+b_2$ and $U_1(t)-b_1\leq
0$. Hence, \Eqref{eq:max} again holds in Case~$2a$.

\noindent
{\bf Case~$2b$}: $U_1(t)-b_1\geq 0$ and $U_2(t)-b_2\geq 0$

In this case, \Eqref{eq:max} becomes
\bequn
U(t)-b\leq U_1(t)-b_1+U_2(t)-b_2\, ,
\eequn
which again holds because of Lemma~\ref{lemma:uw} and the fact that
$b=b_1+b_2$. 

Since the case $U_1(t)-b_1\leq 0$ and $U_2(t)-b_2\leq 0$ is not
possible under Case~$2$ (it would violate Lemma~\ref{lemma:uw}), this
establishes that \Eqref{eq:max} holds in all cases.  Accordingly,
$N(t)\leq N_1(t)+N_2(t)\,, \,\forall t$, so that as mentioned earlier,
$S(t)\leq S^{(2)}(t)\,, \,\forall t$, which establishes the result for
$k=2$.

Extending the result to $k>2$ is readily accomplished by applying the
above approach recursively to groups of two sub-token buckets.
\end{proof}
In concluding, we note that while \Eqref{eq:tb_delay} assumed an FCFS
service ordering for messages in the token bucket, both
Lemma~\ref{lemma:uw} and Proposition~\ref{theo:1vs2} are independent
of the order in which messages waiting for tokens are scheduled for
transmission, as long as the schedule is ``work-conserving,'' \ie
(waiting) messages in any bucket leave as soon as one full token is
available.



\end{document}